\documentclass[preprint]{aastex}

\newcommand{\kms}{\rm{\ km\ s^{-1}}}
\newcommand{\cin}{$C_{in}$ }
\newcommand{\bt}{$B/T$ }
\newcommand{\btn}{$B/T$}
\newcommand{\ewha}{EW(H$\alpha$) }

\shorttitle{The Environmental Dependence in the Local Universe}
\shortauthors{Tanaka et al.}

\begin{document}


\title{The Environmental Dependence of Galaxy Properties in the Local Universe:
Dependence on Luminosity, Local Density, and System Richness}


\author{Masayuki Tanaka\altaffilmark{1}, Tomotsugu Goto\altaffilmark{1,2,3}, Sadanori Okamura\altaffilmark{1},
Kazuhiro Shimasaku\altaffilmark{1}, and Jon Brinkman\altaffilmark{4}}
\email{tanaka@astron.s.u-tokyo.ac.jp, tomo@jhu.edu, okamura@astron.s.u-tokyo.ac.jp, shimasaku@astron.s.u-tokyo.ac.jp, jb@apo.mnsu.edu}
\altaffiltext{1}{Department of Astronomy, School of Science, University of Tokyo,
7-3-1 Hongo, Bunkyo-ku, Tokyo 113-0033, Japan}
\altaffiltext{2}{Institute for Cosmic Ray Research, University of Tokyo,
Kashiwanoha, Kashiwa, Chiba 277-0882 Japan}
\altaffiltext{3}{Department of Physics and Astronomy, Johns Hopkins University,
3400 North Charles Street, Baltimore, MD 21218-2686, USA}
\altaffiltext{4}{Apache Point Observatory, P.O. Box 59, Sunspot, NM 88349, USA}



\begin{abstract}


We investigate the environmental dependence of star formation and the morphology of galaxies
in the local universe based on a volume-limited sample constructed
from the data of the Sloan Digital Sky Survey.
The sample galaxies (19,714 in total) are restricted to the redshift range of $0.030<z<0.065$
and the magnitude range of $M_r < M^*_r+2$.
We investigate correlations between star formation, morphology, luminosity,
local density, and richness of galaxy systems.
First, we focus on how galaxy properties change with local density.
Galaxies in dense environment are found to have suppressed star formation rates
and early morphological types compared with those in the field.
Star formation and morphology show a 'break' at the critical local density of
$\log \Sigma_{\rm crit}\sim 0.4\  \rm galaxies{\it\ h}_{75}^{2}\ \rm Mpc^{-2}$,
which is in agreement with previous studies.
However, the break can be seen only for faint galaxies ($M^*_r+1 < M_r < M^*_r+2$), and
bright galaxies ($M_r < M^*_r+1$) show no break.
Thus, galaxies of different luminosities are found to show different environmental dependencies.
Next, we examine dependencies on richness of galaxy systems.
Median properties of galaxies residing in systems with $\sigma>200\kms$ show no dependence
on system richness, and most of the galaxies in those systems are
non-star-forming early-type galaxies.
Star formation activities of galaxies are different from those of field galaxies even in systems
as poor as $\sigma\sim100\kms$.
This result suggests that environmental mechanisms that are effective
only in rich systems, such as ram-pressure stripping of cold gas and harassment,
have not played a major role in transforming galaxies into red early-type galaxies.
Strangulation and interactions between galaxiesm however, remain candidates of the driver of the
environmental dependence.
In the dense environment in the local universe, the slow transformation of faint galaxies
occurs to some extent, but the transformation of bright galaxies is not clearly visible.
We suggest that the evolution of bright galaxies is not strongly related to galaxy system,
such as groups and clusters, while the evolution of faint galaxies
is likely to be closely connected to galaxy system.
\end{abstract}




\section{INTRODUCTION}


It is widely known that galaxy properties such as
morphology and the rate of star formation depend on environment.
For instance, we know that red early-type galaxies dominate galaxy clusters, whereas 
blue late-type galaxies preferentially reside in the field region
(e.g., \citealt{dressler80,whitmore93}).
Why do galaxy properties vary depending on environment in which they are located?
How is such a relation established in galaxy evolution?
Although considerable effort has been devoted to answer these questions,
no clear conclusion has been reached as yet.

Intensive studies of galaxy clusters have demonstrated a clear connection between
galaxy properties and environment
\citep{dressler80,whitmore93,balogh97,balogh98,hashimoto98,
balogh99,poggianti99,couch01,lewis02,gomez03,blanton03a}
with possible evolution \citep{butcher84,couch87,dressler97,andreon98,fasano00,goto03a,goto03b}.
\citet{lewis02} and \citet{gomez03} statistically analyzed
the star formation rates (SFRs) of galaxies using large-survey data which became available in recent years.
They analyzed a wide range of environment, i.e., from the sparse field to the dense cluster cores,
and showed that star formation is suppressed beyond the virial radius (out to $\sim 2 R_{vir}$)
of clusters with the break at the critical density of $\Sigma \sim 1\ \rm galaxies \rm\ Mpc^{-2}$.
\citet{goto03c} investigated galaxy morphology using the same sample as \citet{gomez03} and
found that morphology also breaks at the same density found for star formation.
\citet{treu03} presented the morphology-density relation in Cl0024 ($z=0.4$) on the basis of 
39 pointings of the WFPC2 on the {\it Hubble Space Telescope}.
They suggested that the morphology-density relation shows no significant evolution in low density regions.
With the wide-field camera Suprime-Cam on the Subaru Telescope \citep{miyazaki02}, \citet{kodama01}
identified galaxy groups surrounding the A851 cluster ($z=0.41$). 
They found a break in galaxy color at a density typical of galaxy groups.
They concluded that star formation is already suppressed in such groups and
that most of the red galaxies may have stopped forming stars before they enter the cluster core.
It had been believed that cluster environment plays a critical role in galaxy evolution, but
\citet{kodama01} illustrated the importance of group environment.

Although galaxy groups have recieved less attention than clusters,
some hints of environmental dependence in groups have been reported.
\citet{postman84} found that the morphology-density relation \citep{dressler80} holds in groups
suggesting the existence of the morphology-density relation over a wide range of local density.
\citet{zabludoff98} and \citet{tran01} studied X-ray selected groups.
Both studies suggested that the properties of galaxies differ from those of field galaxies in the sense
that more red early-type galaxies are found in groups.
\citet{dominguez02} analyzed morphology in groups in the 2dF redshift survey \citep{merchan02}
finding that, in massive groups (${M_{\rm virial}}>10^{13.5}\rm M_\odot$),
the fraction of low SFR galaxies depends on local density and group-centric radius.
On the other hand, no significant dependence was found in less massive groups.
\citet{martinez02} investigated SFRs in galaxy groups and found
a correlation between the relative fraction of star forming galaxies
and the mass of the parent group.

Several physical mechanisms are expected to drive the environmental
dependence of galaxy properties.
Each mechanism has specific environment in which it works most effectively.
For example, ram-pressure stripping \citep{gunn72,abadi99,quilis00} is effective
in the cores of rich clusters, and low-velocity galaxy-galaxy interactions
(e.g., \citealt{mihos94,mihos96,mihos03}) between galaxies are effective in groups.
We roughly categorize the suggested mechanisms into two classes:
mechanisms that are especially effective in rich clusters,
and those that are effective in other environments.
The former class includes, for example,  ram-pressure stripping,
harassment \citep{moore96a,fujita98,moore99}, and interaction with cluster potential \citep{byrd90}.
The latter class includes low-velocity encounters, mergers, and strangulation 
which is often referred to in the literature
as suffocation, strangulation, or halo gas stripping
\citep{larson80,balogh00,diaferio01,okamoto03}.
Therefore, the dependence of star formation and morphology on the richness of galaxy systems is
key to understanding the underlying physical mechanisms.

In this paper, we present a detailed study of the environmental dependence of galaxy properties
on the basis of data from the Sloan Digital Sky Survey (SDSS).
A unique feature of this paper is that we investigate galaxies down to $M_r^*+2$
which is deeper by $\sim 1$ mag than previous volume-limited studies \citep{lewis02,gomez03,balogh04}.
The environmental dependence of faint galaxies ($M_r^*+1<M_r<M_r^*+2$) is addressed in particular in detail.
Another point of emphasis is our statistical analysis of star formation
and galactic morphology from two different points of view.
First, we focus on the density-defined environment and examine relationships between
galaxy properties and environment.
We adopt a definition of density similar to that used in previous studies
(e.g., \citealt{gomez03,balogh04}).
Second, we examine relationships between galaxy properties and richness of galaxy systems
such as groups and clusters to put constraints on the proposed mechanisms.
In \S2, we describe the definition of our sample.
The dependencies of galaxy properties on density-defined environment are investigated in \S3.
The star formation--density and morphology-density relations are also examined in that section.
We focus on the richness of galaxy systems in \S4, and 
discuss the implications of our findings in \S 5.
Finally, our conclusions are summarized in \S 6.

In the course of this work, we have found similar independent work by \citet{balogh04}.
They investigated the environmental dependence of star formation of galaxies in the local universe,
and suggested that the dependence is the product of mechanisms such as galaxy-galaxy interactions
that took place at high redshift.
The reader is referred to the paper for a detailed discussion.

Throughout this paper we assume a flat universe of 
$\Omega_{\rm{M}}=0.3,\ \Omega_{\rm\Lambda}=0.7$, and ${H_0}=75\kms\ Mpc^{-1}$.

\section{SAMPLE DEFINITION}


We use the data from the SDSS (\citealt{york00,stoughton02,abazajian03}).
The plan of the SDSS is to observe 
an unprecedented number of objects both photometrically and spectroscopically in one quarter of the sky.
The imaging survey of the SDSS is made in five optical bands, $u, g, r, i$, and $z$
\citep{fukugita96,hogg01,smith02}.
The reader is referred to \citet{gunn98} for details regarding the SDSS camera system and
to \citet{pier03} for the astrometric calibration.
The spectroscopic part of the SDSS observes essentially all galaxies brighter than
$r=17.77$ selected from the imaging survey (Main Galaxy Sample; Strauss et al. 2002).
With a pair of double fiber-fed spectrographs, 640 spectra are obtained with a 45 minutes
exposure covering from 3800\AA \ to 9200\AA \ with a resolution of $\lambda/\Delta\lambda \sim 2000$.
Each fiber subtends $3''$ on the sky.
Due to the mechanical constraints of the spectrograph, a fiber cannot be located closer than
$55''$ to the nearby fiber, which would otherwise result in ``fiber collision.''
A tiling algorithm has been developed to reduce the number of objects that are not fed fibers
as a result of fiber collision \citep{blanton03b}.
The completeness of the spectroscopic survey is probably somewhere between $90\%$ and $95\%$.

We construct a volume-limited sample based on the SDSS data.
All the photometric data of our sample are extracted from the first data release catalog
(DR1; \citealt{abazajian03}).
We gather spectral data\footnote {All the spectral data analyzed here are spRerun 22.}
from the Data Archive Server
and cross-match them with photometric objects in DR1 to construct a full catalog.
We have not used the spectroscopic objects in DR1 because 
the sky coverage of the DR1 spectroscopic data is very patchy and
does not cover the entire region of the photometric data.
The patchy distribution is a significant disadvantage in the calculation of
galaxy density (see \S 3).
As a check, we cross-matched the objects in our catalog that fall in the DR1 spectroscopic region
with DR1 spectroscopic objects,
and, finding that $\sim99.6\%$ of the objects were matched within a $3''$ separation.
Note that the spectroscopic data of DR1 are constructed from data different from those we use
(e.g., from a different version of the pipeline software used).
Although there are actually some objects in DR1 that are not matched with our objects,
they comprise less than $0.4\%$ of the total.
Thus, the missing objects cause no significant damage to the overall completeness of our sample.
It is worth noting that the missing objects are not biased toward any particular type of galaxy.
That is, there are no such trends as that the missing galaxies tend to be star-forming
galaxies or the missing galaxies tend to be early-type galaxies.

\begin{figure}
\plotone{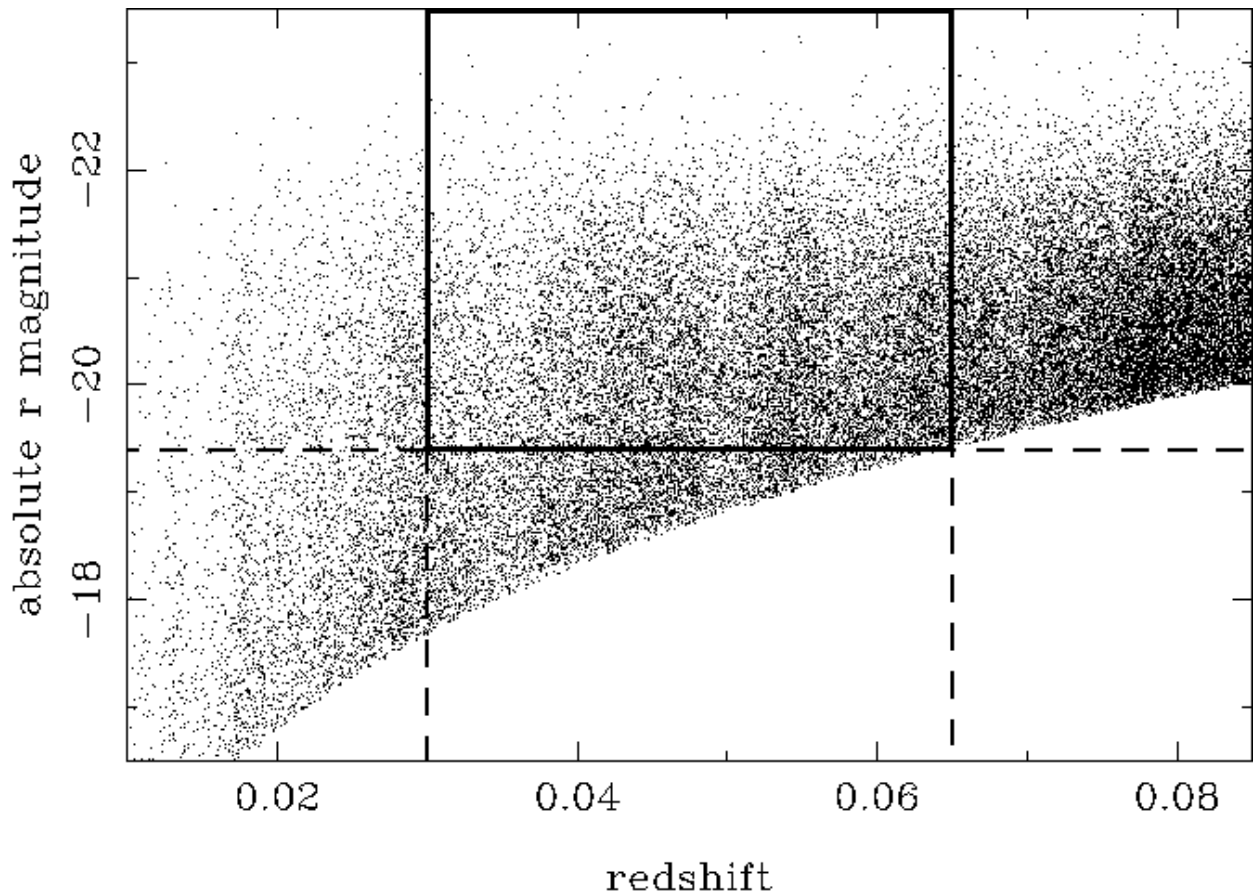}
\caption{
Absolute $r$ magnitude of galaxies in the Main Galaxy Sample \citep{strauss02} plotted against redshift.
We select galaxies in the redshift range of $0.030<z<0.065$ and the magnitude range of
$M_r<M^*_r+2\ (=-19.4)$, i.e., galaxies within the solid rectangle.
\label{fig:volume_limited}
}
\end{figure}

We select galaxies in the redshift range of $0.030 < \rm{z} < 0.065$,
and the magnitude range of $M_r<-19.4$ which corresponds to $M_r<M^*_r+2$
(\citealt{blanton01}; for our cosmology, $M_r^*=-21.4$)
and construct the volume-limited sample as illustrated in Figure \ref{fig:volume_limited}.
Most of the volume-limited samples in previous studies are defined as
$0.05<z<0.1$ and $M_r<M^*_r+1$.
We set the redshift cuts lower than those of the previous work
in order to study fainter galaxies at the cost of reducing the number of sample galaxies.

All the magnitudes in the SDSS are asinh magnitudes \citep{lupton99}.
Since the difference between asinh magnitudes and conventional logarithmic magnitudes is
negligible in our sample,
we apply no conversion and treat asinh magnitudes as standard AB magnitudes \citep{fukugita96}.
Galactic extinction is corrected for each band \citep{schlegel98}.
A K-correction is applied to all magnitudes using the code by \citet[v1\_16]{blanton03c}.
The estimated values of the k-correction are small, e.g., $\Delta m_{r}\sim 0.05$, and are
in agreement with those calculated using \citet{fukugita95}.
Whether or not we apply the k-correction does not change our results.
Also note that the errors in the photometry and the k-correction have no effect on our results.

In summary, our volume-limited sample contains $19,714$ objects brighter than
$M^*_r+2\ (=-19.4)$ over the redshift range of $0.030<z<0.065$.\\

\section{DEPENDENCE ON LOCAL DENSITY}


In this section, we examine the dependence of star formation and morphology on density-defined environment.
We characterize environment by a surface galaxy density defined in the next subsection.
Statistical analyses of density dependencies have been performed by several authors
(e.g., \citealt{lewis02,gomez03,goto03c,balogh04}),
but their samples are restricted to galaxies with $\lesssim M_r^*+1$.
Our sample reaches $M_r^*+2$, and we focus on how the properties of galaxies of different luminosities depend
on environment.

\subsection{Definition of Surface Galaxy Density}
We determine surface galaxy density from the distance to the fifth
nearest neighbor in a pseudo three--dimensional fashion.
Since the galaxies in our sample have spectroscopic redshifts,
we define a redshift sheet of $\pm 1000\kms$ around the galaxy in question.
All galaxies within the sheet are projected onto the redshift of the galaxy in question.
Galaxies outside the  $\pm 1000\kms$ sheet are not used in the calculation.
Then we search for the fifth nearest neighbor and calculate surface galaxy density
in units of galaxies per square megaparsec.
This is exactly the same definition as the one adopted by \citet{miller03}, \citet{goto03c}, and \citet{balogh04}.
Following the convention, we term the density defined in this way ``local density.''
Note that local density is actually a {\it surface density}.
Also note that the median error in recession velocity estimates in our sample is $\sim 20\kms$,
and we neglect effects of the error on density estimates.

All galaxies that reach the survey boundary before finding their fifth
nearest neighbor are excluded from the following analysis because the local densities of such
galaxies are not correctly calculated.
They are, however, used in the local density calculation of other galaxies.
As for galaxies close to the redshift cuts, we make a small compromise not
to exclude all the galaxies to gain sample statistics.
We exclude all galaxies that lie within $500\kms$ from the redshift cuts, but use
galaxies between $500$ and $1000\kms$ from the redshift cuts by applying a volume correction.
Therefore, all galaxies have a sheet width of at least $1500\kms$, so the maximum correction factor is
$2000/1500=1.33$.
We have confirmed that whether we make the compromise or not does not alter our conclusions.
Because of  fiber collision, some galaxies in dense environment are not spectroscopically observed,
and we tend to underestimate densities in dense environment.
We find, however, that our results are not strongly affected by the fiber collision.
A brief discussion on this point is given in Appendix A.

In our catalog, $12,376$ galaxies out of $19,714$ have local density estimates
and the rest of the galaxies are too close to the boundary.

\subsection{The Star Formation-Density Relation}
We investigate a correlation between star formation and local density.
To evaluate star formation in galaxies, we use the $g-i$ color
and the equivalent width (EW) of the $\rm H\alpha$ emission line.

Since we define our volume-limited sample at relatively low redshifts, a large fraction of 
galaxy light is missed by the $3''$ diameter fiber.
It is thus expected that a significant fraction of the fluxes of nebular emission lines
is not observed, since such lines mostly originate from galaxy disks rather than bulges.
Accordingly, we tend to underestimate the flux of the emission lines.
Therefore, care is needed when interpreting emission--line properties in our sample.
A brief discussion of the aperture bias is given in Appendix B.
In the main text of this paper, we focus on the relative strength, rather than the absolute strength,
of star formation activity using EW(H$\alpha$).
We do this because errors in the absolute star formation rate estimates are difficult to estimate,
so we make various corrections to derive absolute star formation rates from observed spectra,
and the uncertainty in each correction is not well known.
Nevertheless, even with such difficulties and uncertainties,
the environmental dependence of absolute star formation activity is worth investigating,
and we present this analysis in Appendix C.
We also examine the $g-i$ color, which is measured for the entire galaxy
and free from such fixed-aperture bias.
Note that we apply no correction for the internal extinction of EW(H$\alpha$).

Galaxies showing an AGN signature are not used when analyzing EW(H$\alpha$).
We adopt ${\rm [NII]/H\alpha}$ vs. ${\rm [OIII]/H\beta}$ diagnostics to select AGNs.
AGNs are defined as the $1\sigma$ lower limit of the model defined by \citet{kewly01}.
When some of the lines are not available, we make use of the two-line method \citep{miller03}.
In our sample, $3371$ galaxies are classified as AGNs.
Whether or not we remove AGNs does not affect our results.

In the following analysis, we classify galaxies into two subclasses by absolute magnitude.
Hereafter, galaxies with $M_r<M_r^*+1$ are referred to as {\it bright galaxies},
and those with $M_r^*+1<M_r<M_r^*+2$ as {\it faint galaxies}.
Our definition of the subsamples and the number of galaxies in each subsample are summarized in Table 1.

\begin{figure}
\plotone{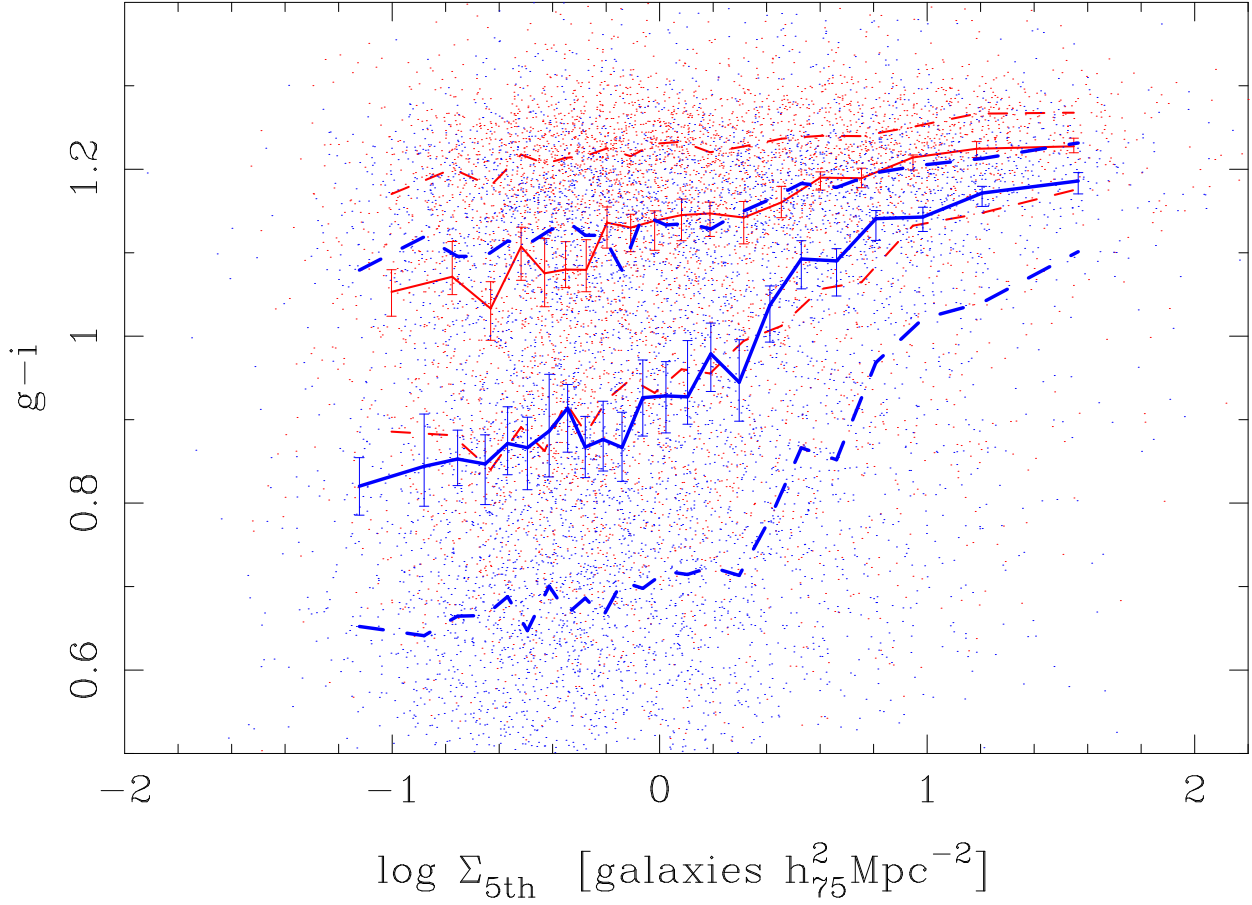}
\caption{
Plot of $g-i$ against local density.
The red and blue lines represent bright and faint galaxies, respectively
(see Table 1 for the definition of the subsamples).
The solid and dashed lines show the median and the quartiles (25\% and 75\%) of the distribution.
The median lines are accompanied by the 90th percentile interval bars estimated by the bootstrap
resampling including the measurement error in $g-i$.
Each bin contains 300 galaxies.
\label{fig:local_vs_color}
}
\end{figure}

Figure \ref{fig:local_vs_color} shows a correlation between $g-i$  and local density.
The error bars represent the 90th percentile intervals estimated by bootstrap resampling.
The error includes the measurement error in $g-i$, assuming that
the measurement error follows a Gaussian distribution.
A clear trend can be seen in the sense that galaxies become redder as the environment
in which they are located becomes denser.
Bright galaxies show a correlation over the entire range of local density and
the color is a smooth function of local density.
In contrast, faint galaxies have a break at a density of
$\log \Sigma_{\rm 5th}\sim 0.4\ \rm galaxies{\it\ h}_{75}^{2}\ \rm Mpc^{-2}$.
Above the break, galaxies abruptly become redder, while below the break,
the color gradually changes with local density.
Faint galaxies are generally bluer than bright galaxies,
and the trend is particularly prominent in low density regions.
On the other hand, the trend is largely reduced in dense environments, where
faint galaxies are only slightly bluer than bright galaxies.
Note that the small color difference in very dense regions ($\log\Sigma_{\rm 5th}>1$)
reflects the color-magnitude relation (e.g., \citealt{bower92,kodama97}).

\begin{figure}
\plotone{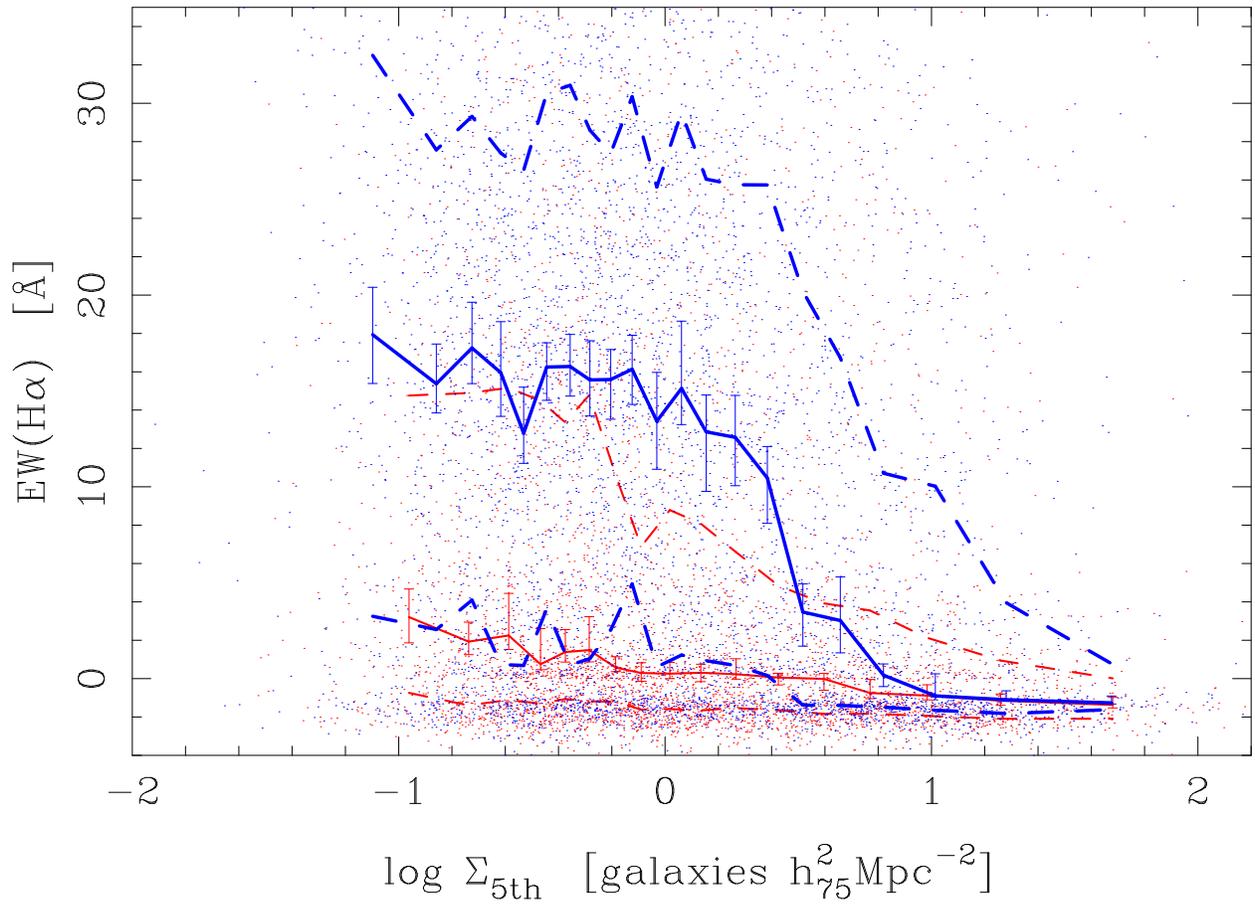}
\caption{
EW of H$\alpha$ plotted against local density.
The meanings of the lines are the same as in Figure 2.
\label{fig:local_vs_ha}
}
\end{figure}

A correlation between \ewha and local density is shown in Figure \ref{fig:local_vs_ha}.
On-going star formation in galaxies measured from \ewha also shows a dependence on local density.
The median \ewha of bright galaxies shows only a little change with local density.
But, the $75^{th}$ percentile shows a monotonic decrease with increasing local density,
and no clear break can be seen.
Faint galaxies have a strong break at the same density as found for $g-i$ (Fig. \ref{fig:local_vs_color}).
The change in median \ewha against local density is larger for faint galaxies.
On average, faint galaxies are more actively forming stars than bright galaxies.
This trend is particularly noticeable in low-density regions.
On the other hand, the star formation activity of faint galaxies is largely suppressed in dense regions,
where most galaxies are quiescent independent of luminosity.
As discussed in Appendix B, the aperture bias in \ewha cannot be ignored,
and if the trends in \ewha are caused by the aperture bias is a major concern.
However, the overall trend seen for \ewha is similar to that seen for $g-i$,
which is free from the fiber aperture bias.
This suggests that the observed trends in \ewha are not a product of the aperture bias.

Let us investigate the observed trend further.
Following \citet{balogh04}, we define galaxies having EW(H$\alpha)>4\rm\AA$ as star-forming galaxies
and examine their EW(H$\alpha$).
Although the aperture bias in our sample is stronger than that of \citet{balogh04},
our discussion does not strongly rely on a particular choice, within a reasonable range,
of the threshold value of \ewha.
This investigation is particularly interesting because star-forming galaxies are expected to show
reduced star formation rates if they are affected by environmental mechanisms that trigger
slow truncation of star formation activities (e.g., strangulation).
If galaxy star formation is suppressed on a very short time-scale (e.g., by mergers),
we do not expect to see strong changes in the \ewha distribution of star-forming galaxies
(the fraction of star-forming galaxies relative to non-star-forming galaxies will change).
Here we aim to see whether or not there is a signature of the slow truncation.

\begin{figure}
\plotone{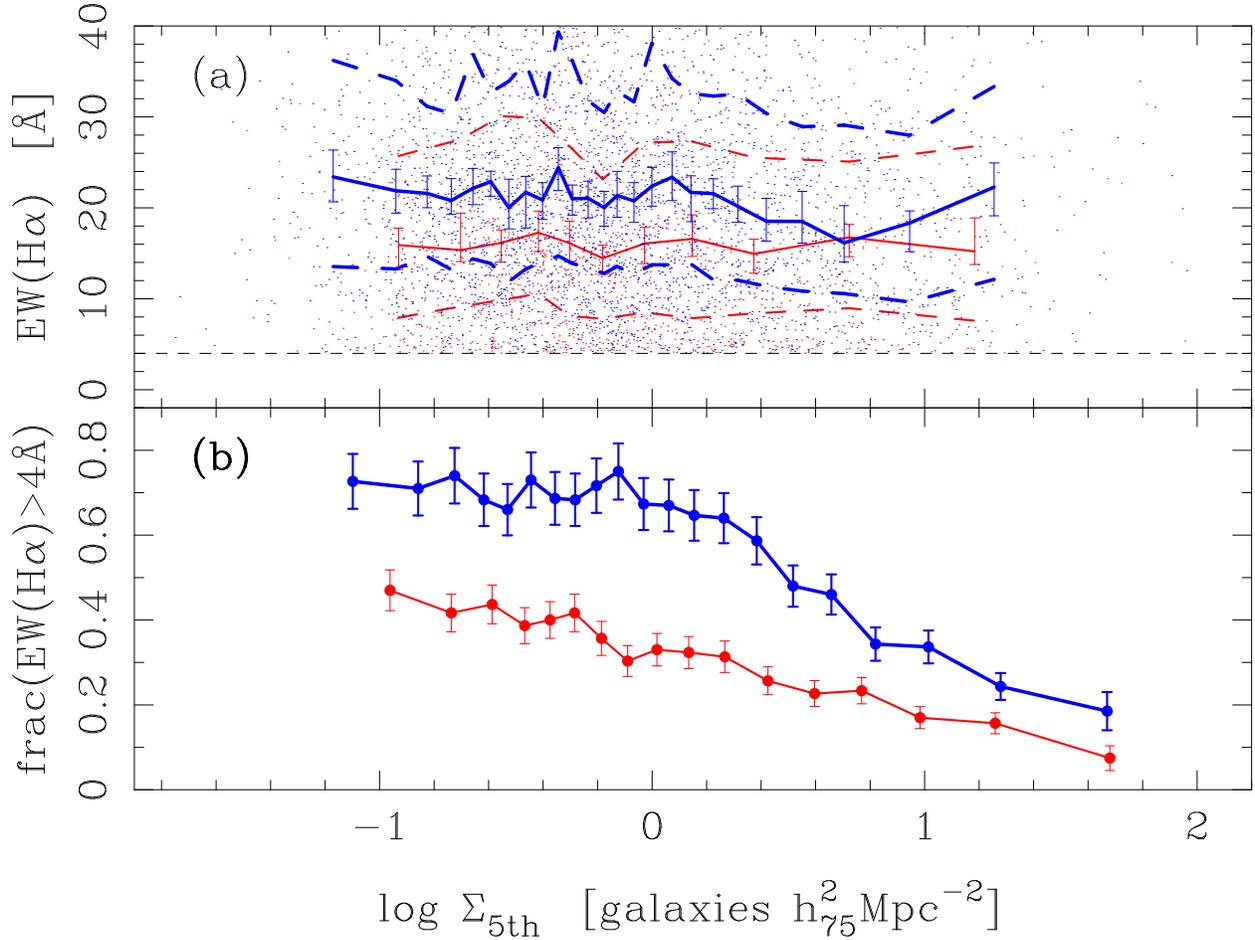}
\caption{
{\bf Top panel (a):} Same as Figure \ref{fig:local_vs_ha}, but galaxies are restricted to those having EW(H$\alpha)>4\rm\AA$ here.
{\bf Bottom panel (b):} Fraction of EW(H$\alpha)>4\rm\AA$ galaxies plotted against local density.
The red and blue lines represent bright and faint galaxies, respectively.
The error bars show $1\sigma$ errors based on the Poisson statistics.
\label{fig:local_vs_ha4}
}
\end{figure}

In the top panel of Figure \ref{fig:local_vs_ha4}, we show the \ewha of star-forming galaxies
as a function of local density.
Bright star-forming galaxies do not show any dependence on local density.
A Kolmogorov-Smirnov (K-S) test does not reject the hypothesis that galaxies above and below the critical density
($\log \Sigma_{\rm 5th}\sim0.4$) are drawn from the same parent population
(a K-S probability of $\sim 60\%$).
The \ewha of faint star-forming galaxies becomes slightly smaller above the critical density.
The K-S test shows that faint star-forming galaxies above and below the critical
density are different ($<0.001\%$).
This should be considered to be a signature of the slow truncation of
the star formation activity of faint galaxies.
We will further discuss this finding below.

The bottom panel of Figure \ref{fig:local_vs_ha4} shows the fraction of star-forming galaxies
as a function of local density.
It can be seen that the fraction of star-forming galaxies decreases strongly with increasing local density.
Combining the bottom panel with the top panel shows
that the trend of bright galaxies in Figure \ref{fig:local_vs_ha}
entirely reflects the change in the fraction of star-forming galaxies,
rather than the change in the star formation activity of star-forming galaxies.
As for faint galaxies, the critical density is caused by the change
in the fraction of star-forming galaxies and the change in the star formation activity
of star-forming galaxies.
However, the former seems to be a stronger effect.

\subsection{The Morphology-Density Relation}
Next, we examine how the morphology of galaxies changes with local environment.
We adopt the bulge--to--total luminosity ratio (\btn) based on
the growth curve fitting technique \citep{okamura99} as an indicator of galaxy morphology.
We briefly summarize the basic idea of the growth curve fitting method below.
The reader is referred to \citet{okamura99} for further details.

We prepare PSF-convolved template profiles of various model galaxies
that have a $r^{1/4}$ bulge and an exponential disk.
There are three parameters in the model; the ratio of the effective radii of the bulge and the disk,
the ratio of the effective radius of the disk to the size of seeing,
and the ratio of the bulge to the total luminosity (\btn).
The template profiles are generated for a combination of the parameters in reasonable ranges.
Then we compare the profile of a real galaxy with the template profiles and find the best--fitting template.
We make use of the \bt of the best--fitting template as the indicator of morphology. 
Note that the SDSS measures the profile of a galaxy in a sequence of circular apertures, 
and so the inclination of galaxies may affect \bt estimates.
Using only low--inclination galaxies ($b/a>0.7$), we have confirmed
that our results are not affected by the inclination effect.
Note as well that each model galaxy is assumed to be composed of a $r^{1/4}$ bulge and a exponential disk.
However, of course, real galaxies have more or less different radial profiles for each component.
The statistical robustness of this model has been confirmed
by examining a correlation between visually classified
morphology and the \bt of the best fitting template for nearby galaxies (\citealt{okamura99}; see their Figure 11).

\begin{figure}
\plotone{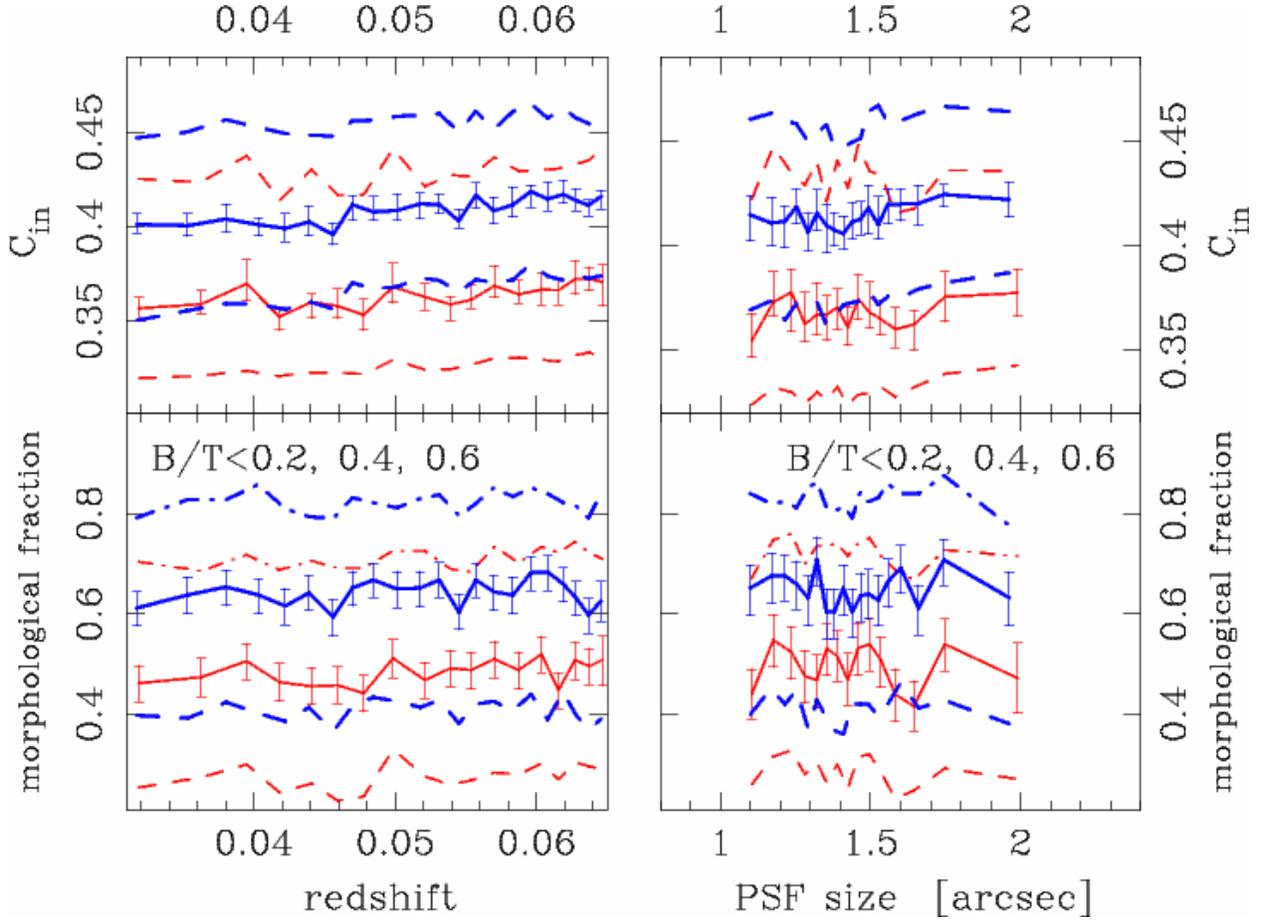}
\caption{
\label{fig:seeing_effect}
{\bf Left panels:} Morphology (\cin and \bt) plotted against redshift.
The red and blue lines represent bright and faint galaxies, respectively.
In the top panels, the lines show the median and the quartiles of the distribution.
In the bottom panels, the lines show the fraction of $B/T<0.2$ (dashed), $B/T<0.4$ (solid),
and $B/T<0.6$ (dot-dashed) galaxies.
The error bars are the bootstrap $90\%$ intervals.
{\bf Right panels:} Morphology (\cin and \bt) plotted against the PSF size.
Galaxies are restricted to the redshift range of $0.055<z<0.065$ to minimize
the redshift dependence (left panel).
}
\end{figure}

We demonstrate that our \bt estimates are robust against the seeing effect.
We present \bt along with the inverse concentration index (\cin; \citealt{shimasaku01};
see also \citealt{strateva01}) in Figure \ref{fig:seeing_effect}.
Note that \cin is not corrected for seeing.
The seeing effect is investigated from two different points of view.
First, we examine \bt and \cin and as a function of redshift.
Since galaxies look smaller at higher redshift, galaxies at higher redshift are
more severely blurred by the seeing.
This redshift effect is shown in the left panels.
It can be seen that \cin increases ($\Delta C_{in}\sim0.02$)
with increasing redshift, while \bt remains essentially unchanged.
The right panels show \bt and \cin as a function of the PSF size.
To minimize the redshift effect described above, we select galaxies within $0.055<z<0.065$.
\cin increases slightly with increasing the PSF size,
while \bt does not show any significant change.
From these results, we conclude that our \bt estimates are robust against the seeing effect.
Since faint galaxies are smaller than bright galaxies and they are more easily affected by
the seeing effect, one may suspect that morphologies of faint galaxies are biased.
But, we expect this is not the case.

\begin{figure}
\plotone{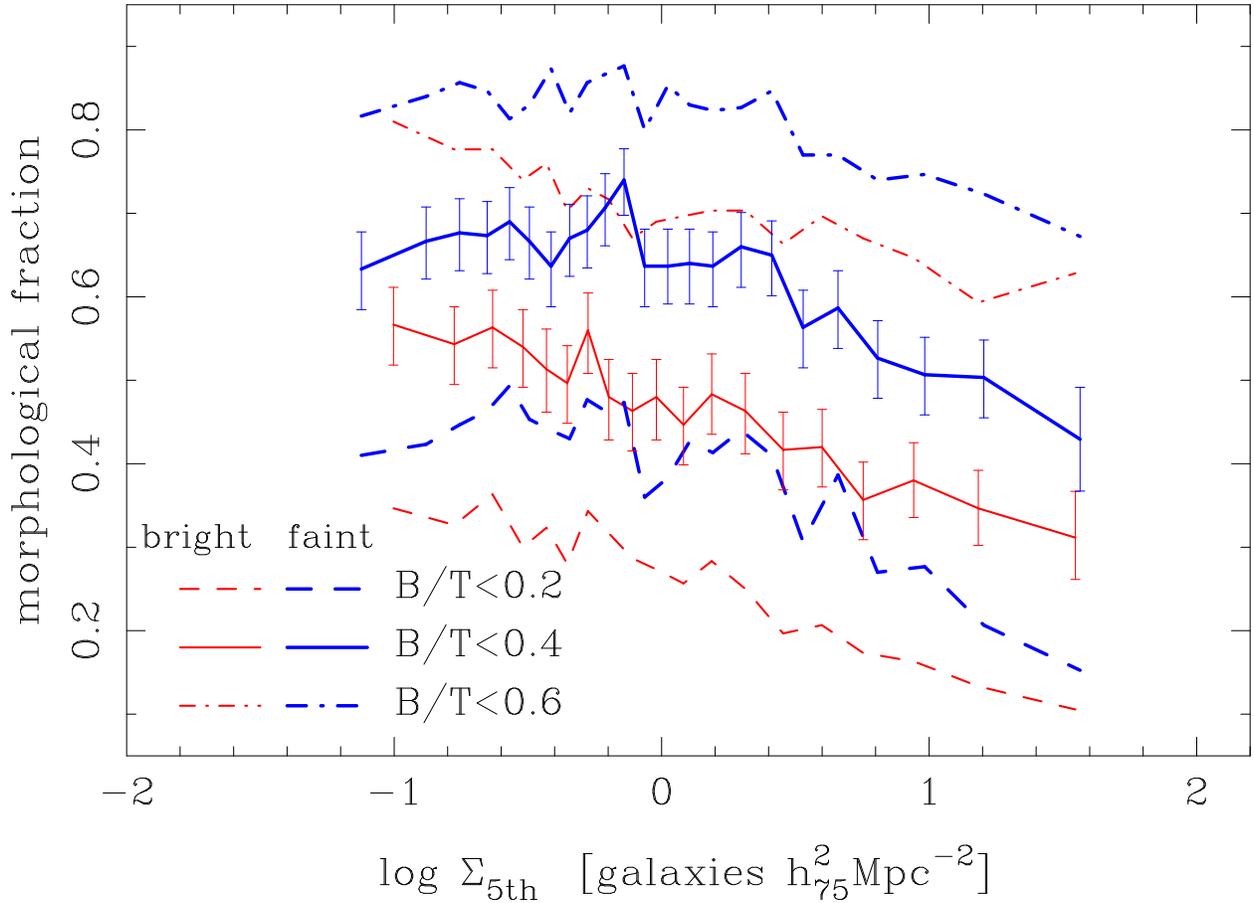}
\caption{
Morphological fractions determined by \bt plotted against local density.
The red and blue lines represent bright and faint galaxies, respectively.
Each bin contains 300 galaxies, and the lines show the fraction of
$B/T<0.2$ (dashed), $B/T<0.4$ (solid), and $B/T<0.6$ (dot-dashed) galaxies in each bin.
The error bars are the bootstrap $90\%$ intervals.
Note that a measurement error in \bt is not included in the error bars here.
\label{fig:local_vs_bt}
}
\end{figure}

Figure \ref{fig:local_vs_bt} shows the morphological fractions based on \bt as a function of local density.
Because of the coarse grid of \bt measurements ($\Delta B/T=0.1$), we cannot take the median of \btn.
We adopt threshold values of $B/T=0.2,\ 0.4$, and $0.6$ to analyze the morphological mix.
An measurement error in \bt is difficult to quantify
since \bt is estimated via multi-parameter template fitting.
Therefore an error in \bt is not considered here.
The fraction of late-type ($B/T< 0.2,\ 0.4,\ \rm and\ 0.6$)
bright galaxies decreases monotonically with increasing local density,
and no break can be seen.
Faint galaxies show very different behavior, with a visible break
at $\log \Sigma_{\rm crit} \sim 0.4$.
The late-type fractions do not show any significant change below this critical density,
whereas a strong change can be found above the critical density, just like the trend seen in star formation.
It seems that the morphological change of bright galaxies with local density is
stronger than or as strong as that of faint galaxies.
This is in contrast to the trend found for star formation
[for which the changes in $g-i$ and \ewha are stronger for faint galaxies].
Moreover, it is interesting to note that faint galaxies are morphologically later than bright galaxies
at any local density.
Because morphology is somewhat ambiguous to quantify, it is worth using \cin as another indicator, in addition to \btn, although \cin is not stable in terms of seeing.
We have found that \cin shows almost identical environmental trends with \btn, and hence
we do not present a \cin plot here.

In the previous subsection, we showed how the \ewha of star-forming galaxies [EW(H$\alpha)>4\rm\AA$]
changes with local density (Fig. \ref{fig:local_vs_ha4}).
We now perform a similar analysis for morphology.
For this purpose, we use \cin instead of \bt because of the coarse grid of the \bt estimates.
We examine galaxies with $C_{in}>0.45$ (late-types) to see whether there is a slow transformation of morphology
in late-type galaxies.
We find that, based on the K-S test, the morphological distribution of bright late-type galaxies
is not significantly different above and below the critical density (the K-S probability of $\sim30\%$),
while that of faint late-types may be different ($\sim5\%$)
and they may have systematically smaller \cin above the critical density.

\subsection{The Bimodality of Galaxy Properties}
It is known that the distribution of star formation and morphology is bimodal
\citep{strateva01,kauffmann03,blanton03a,baldry04}
--- namely, there are two distinct populations of galaxies,
star-forming late-type and non--star-forming early-type galaxies.
This bimodality is observed up to $z\sim1$ \citep{bell03a,bell03b}.

\begin{figure}
\plotone{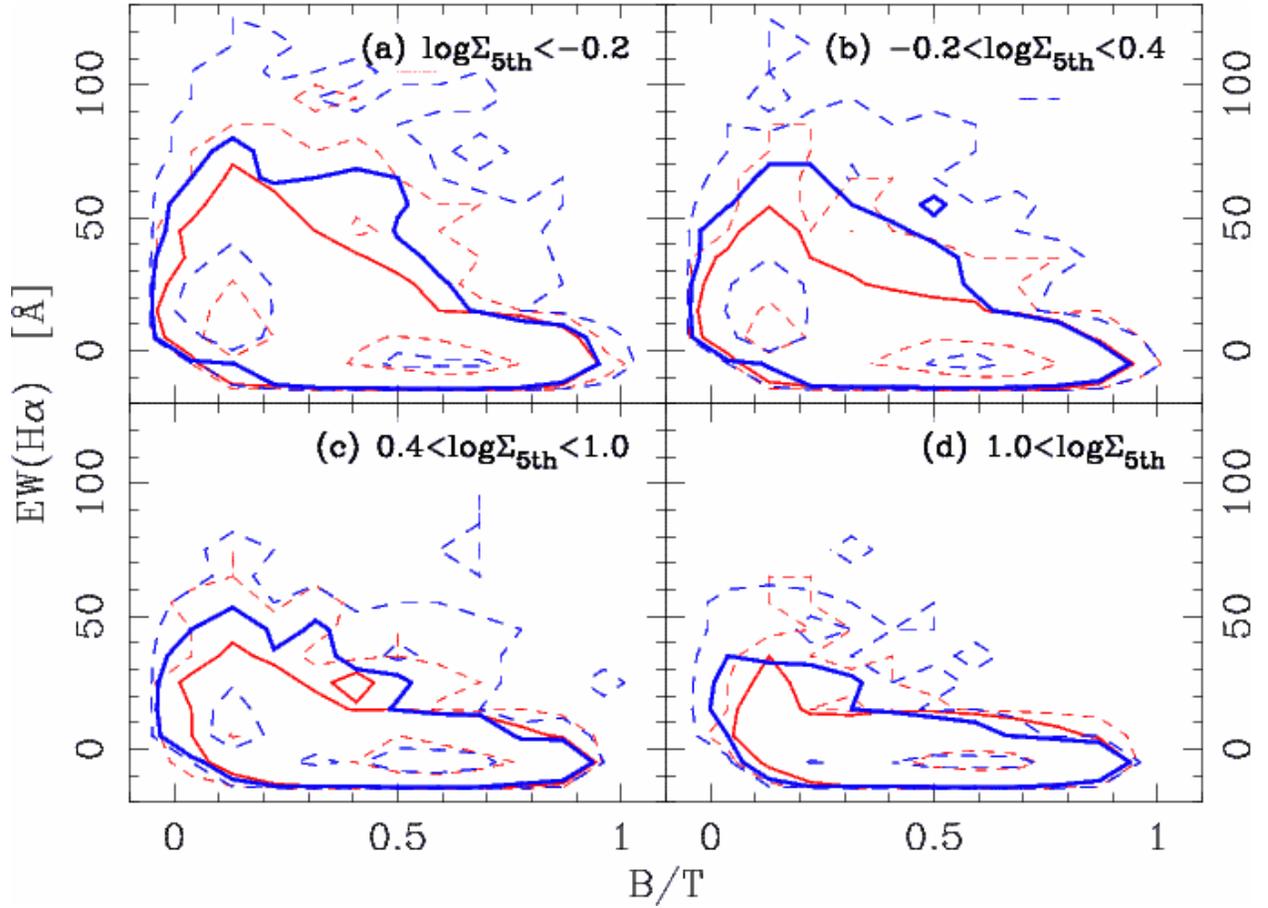}
\caption{
Distribution of galaxies in the \bt vs. \ewha plane for four different local density ranges:
{\bf (a)} $\log\Sigma_{\rm 5th}<-0.2$, {\bf (b)} $-0.2<\log\Sigma_{\rm 5th}<0.4$,
{\bf (c)} $0.4<\log\Sigma_{\rm 5th}<1.0$, and {\bf (d)} $1.0<\log\Sigma_{\rm 5th}$.
The red and blue lines represent bright galaxies and faint galaxies, respectively.
The solid lines show the median of the distribution.
The dashed lines show the $10^{th}$ and $90^{th}$ percentiles of the distribution.
\label{fig:bt_vs_ha}
}
\end{figure}

To sum up the analyses of previous subsections, we present in Figure \ref{fig:bt_vs_ha}
the relationship between star formation and morphology in the four local density ranges.
Faint galaxies show more active star formation than bright galaxies at a given \btn.
This trend is particularly prominent in low--density regions \citep{hogg03a}.
Thus, the relationship between star formation and morphology depends on both luminosity and environment.
However, what strongly changes with environment is
the fraction of non--star-forming early-type galaxies relative to star-forming late-types.
Further illustrations of correlations between star formation, morphology, luminosity, and local density
are presented in Appendix D.

\subsection{The Brightest Galaxies}
\begin{figure}
\plotone{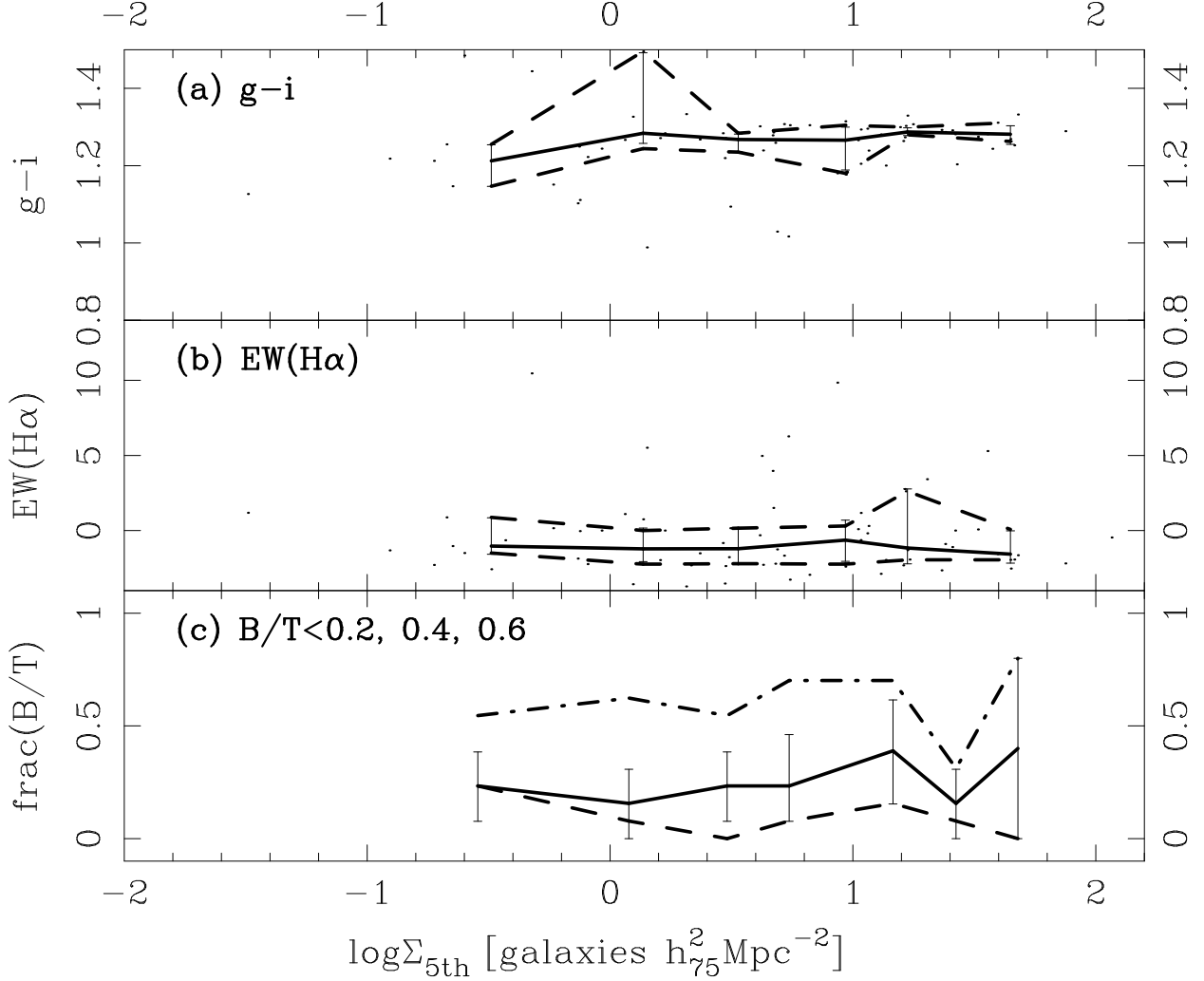}
\caption{
Local density dependence of star formation and morphology of the brightest galaxies ($M_r<M_r^*-1$). 
{\bf(a)} $g-i$, {\bf (b)} EW(H$\alpha$), and {\bf (c)} \btn.
Each bin contains only $\sim 10$ galaxies.
The solid and dashed lines in (a) and (b) represent the median and the quartiles of the distribution.
The lines in (c) are the late-type fractions of $B/T<$0.2, 0.4, 0.6.
The error bars show the bootstrap $90\%$ intervals.
\label{fig:brightest_gals}
}
\end{figure}

We briefly address the local density dependence of star formation and morphology of
the brightest galaxies ($M_r<M_r^*-1$).
As shown in Figure \ref{fig:brightest_gals}, the star formation and morphology of the brightest galaxies show
little dependence on local density.
The brightest galaxies are all red early-type galaxies regardless of local environment.
We have shown that galaxies of different luminosities have different local density dependencies.
The brightest galaxies lie on the extreme end of that trend; they have no dependence on local density.
It should be noted that these brightest galaxies are preferentially found in high density regions
\citep{hogg03b}.  In that sense, their presence itself depends on environment.

\subsection{Summary of the Dependencies on Local Density}

In this section we summarize how star formation and morphology change with local density and luminosity.
We find that star formation is strongly correlated both with local density and luminosity.
Bright galaxies show a monotonic change in $g-i$ and \ewha with local density.
Faint galaxies have a remarkable break at
$\log \Sigma_{\rm crit} \sim 0.4\ \rm galaxies{\it\ h}_{75}^{2}\ \rm Mpc^{-2}$.
The changes in $g-i$ and \ewha from the lowest to highest density are
stronger for faint galaxies.
The changes primarily reflect the change in the population fraction of star-forming galaxies
relative to non--star-forming galaxies.
The luminosity dependence is particularly noticeable in the field, in the sense that
faint galaxies are more actively forming stars.
However, this trend is largely reduced in dense environment, where
most galaxies are not forming stars regardless of luminosity.

Morphology is also a strong function of both local density and luminosity.
A break is found at the same critical density as that found for star formation, again only for faint galaxies.
The morphological change of bright galaxies with local density is stronger than or as strong as
that of faint galaxies.
This is in contrast to the trend found for star formation, implying that 
morphology depends on luminosity in a different way than star formation.
For example, in very dense regions ($\log\Sigma_{\rm 5th}>1$),
star formation is extremely weak regardless of luminosity,
but morphology varies as a function of luminosity.

We have shown that star formation and morphology show slightly different behaviors
depending on local density and luminosity.
This is partly because star formation and morphology are independent properties
\citep{balogh98,hashimoto98,lewis02,gomez03}.
The bimodality of galaxy properties is clearly seen in our sample.
We find that what strongly changes with local density is the fraction of blue late-type galaxies
relative to red early-type galaxies, i.e., the relative heights of the two peaks in the bimodal
distribution change.
Further illustrations of the environmental dependencies are presented in Appendix D.

Particularly interesting trends can be summarized as follows (see also Table 2):

\begin{enumerate}
\item For faint galaxies, there is a critical density of
$\log \Sigma_{\rm crit} \sim 0.4\ \rm galaxies{\it\ h}_{75}^{2}\rm\ Mpc^{-2}$
where star formation and morphology change drastically.
\item No such critical density is found for bright galaxies.
\item The star formation and morphology of the brightest galaxies ($M_r<M_r^*-1$)
do not correlate with local density, and they are all red early-type galaxies
regardless of environment.
\item The population fraction of blue late-type galaxies relative to red early-type galaxies
varies strongly with local density.
\item The star formation activity of bright star-forming galaxies [EW(H$\alpha)>4\rm\AA$] does not
depend on local density, while that of faint star-forming galaxies is systematically weaker above
the critical density. A similar trend seems to be apparent for morphology.
\item The morphological change of bright galaxies with local density is stronger than, or as strong as,
that of faint galaxies, whereas the change in star formation is stronger for faint galaxies than for bright galaxies.
\end{enumerate}

\section{DEPENDENCIES ON SYSTEM RICHNESS}


In this section, we focus on galaxy systems such as groups and clusters and examine
galaxy properties as a function of system richness.
This investigation is particularly interesting since galaxies in poor and rich systems are
considered to be in different stages  of structure formation.
Accordingly, the amount of environmental effects they have suffered should be different.
Previous studies have reported correlations between galaxy properties and system richness.
For example, the fraction of spiral galaxies decreases as systems become rich
(e.g., \citealt{edge91,balogh02,goto03a}),
although some poor systems have an early-type fraction identical to that of rich systems \citep{zabludoff98}.
Moreover, star formation activity may be a function of system richness
(\citealt{margoniner01,martinez02,goto03a}; but see also \citealt{fairley02}).

We apply a friends-of-friends algorithm (FOFA) to find galaxy systems in our sample.
For each galaxy system, we measure velocity dispersion and use it as an indicator of system richness.
Based on the extensive, high-quality data of the SDSS, we examine galaxies in galaxy systems in detail.
To avoid duplicating discussion, we use only \ewha as an indicator of star formation in this section.
We have confirmed that $g-i$ follows a similar trend to \ewha.

\subsection{The Friends-of-Friends Algorithm}
The FOFA is a famous group finding algorithm
originally proposed by \citet{huchra82}.
The statistical properties of the algorithm has been intensively studied
using mock catalogs of N-body simulations and  is well known
(\citealt{moore93,frederic95,ramella97,diaferio99,merchan02}).

The basic strategy of the FOFA is to find a chunk of galaxies connected within certain threshold lengths.
The threshold length is called the linking length.
There are two linking lengths:  angular separation, $D_0$, and line-of-sight velocity difference, $V_0$.
There is one more parameter in the FOFA, $N_{min}$, which is a threshold of the number of member galaxies.
When a resultant chunk has galaxies equal to or greater than $N_{min}$,
the chunk is defined as a galaxy system.

Previous studies have shown that the contamination of chance interlopers in FOFA groups is
relatively small with a density contrast
of $\Delta\rho/\rho>80$ \citep{ramella97,diaferio99,merchan02}. The density contrast is defined as
\begin{equation}
\frac{\Delta\rho}{\rho}=\frac{3}{4\pi D_0^3} \left ( \int^{ M_{lim}}_{-\infty}\Phi  (M) dM \right )^{-1}-1,
\end{equation}
where $\Phi (M)$ is a luminosity function and $M_{lim}$ is a limiting absolute magnitude.
We estimate the value of $D_0$ with the luminosity function
given in \citet{blanton01}. Adopting $\Delta\rho/\rho\approx 160$,
we obtain $D_0=500{\it\ h}_{75}^{-1}\rm\ kpc$.
The statistical properties of the resulting groups are not sensitive to the choice of $D_0$
\citep{frederic95,ramella97,diaferio99,merchan02}.
On the other hand, $V_0$ is the key parameter in constructing a group catalog.
After performing some experiments, we found that $V_0=500\kms$ is a fairly reasonable choice
and we adopt this value in our analysis.
By changing the value of $V_0$ from $300\kms$ to $700\kms$ with a step of $100\kms$,
we have confirmed that our results do not change significantly.
The final parameter $N_{min}$ is set at 5 since a large fraction of systems
with only a few galaxies is expected to be spurious \citep{frederic95,ramella97,ramella02}.
It should be noted that we run the FOFA in the volume-limited sample,
which means the linking lengths are fixed independently of redshift.
We note that the linking lengths are elongated along the line of sight (0.5 Mpc on the sky
and 6.7 Mpc along the line-of-sight) to account for the ``finger--of--God'' effect.

We calculate the velocity dispersions of
galaxy systems using the gapper method \citep{beers90}.
We found that the results presented in the next subsection are basically unchanged if we implement
the biweight estimator instead of the gapper method.

\subsection{Results}
We find 307 galaxy systems in our volume-limited sample.
Systems that lie too close to the survey boundary and redshift cuts are not included.
To be specific, if one of the member galaxies is closer than $\sim 1'$ to the survey boundary
or if the redshift cuts lie closer than $2\sigma$ of velocity dispersion the system will not
be used in the following analysis.
As shown in Appendix E, galaxies in our FOFA catalog are not confined to the system core,
but extend to the outskirts.  Our system is a fair representation of a system as a whole.

\begin{figure}
\plottwo{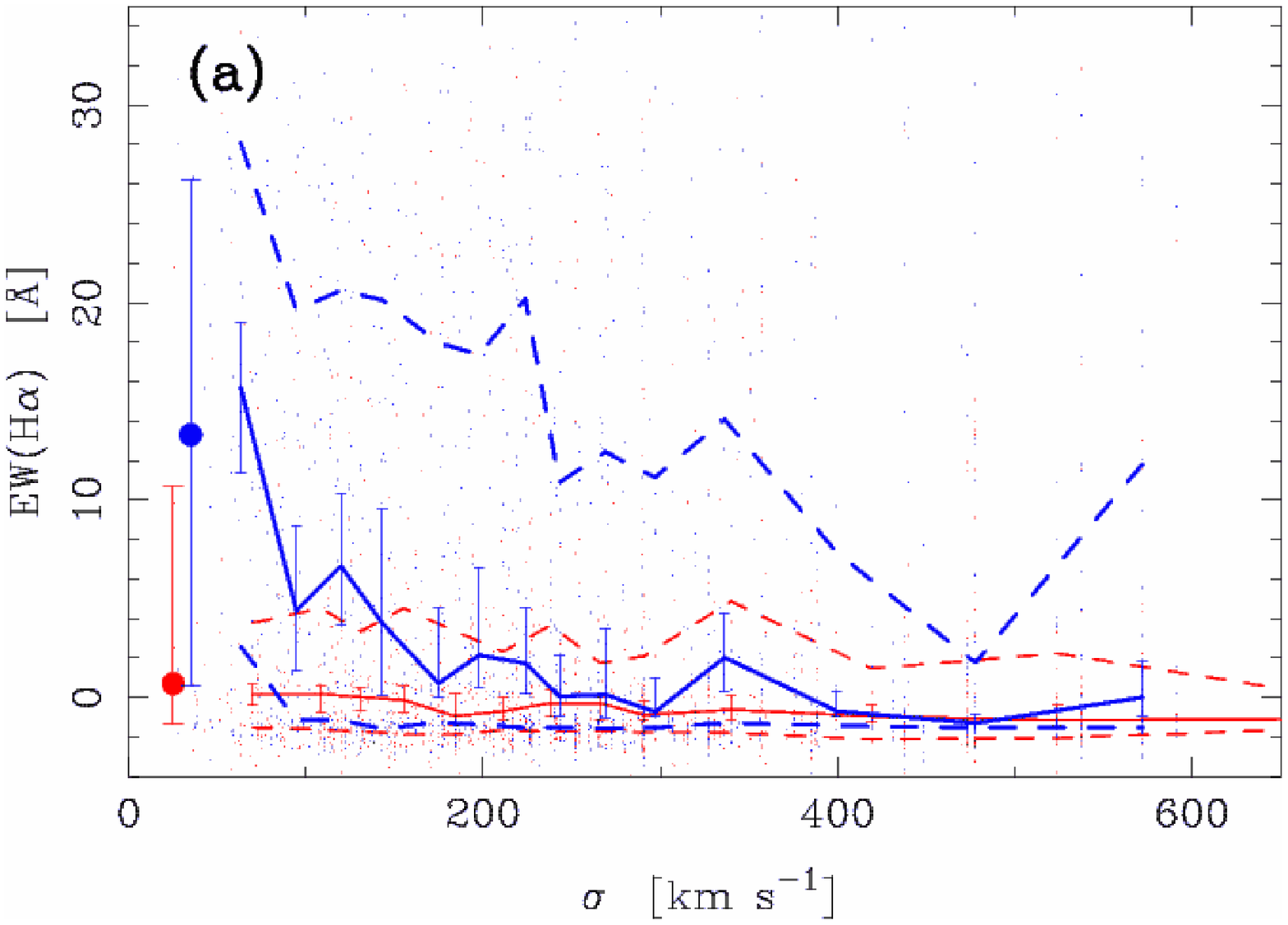}{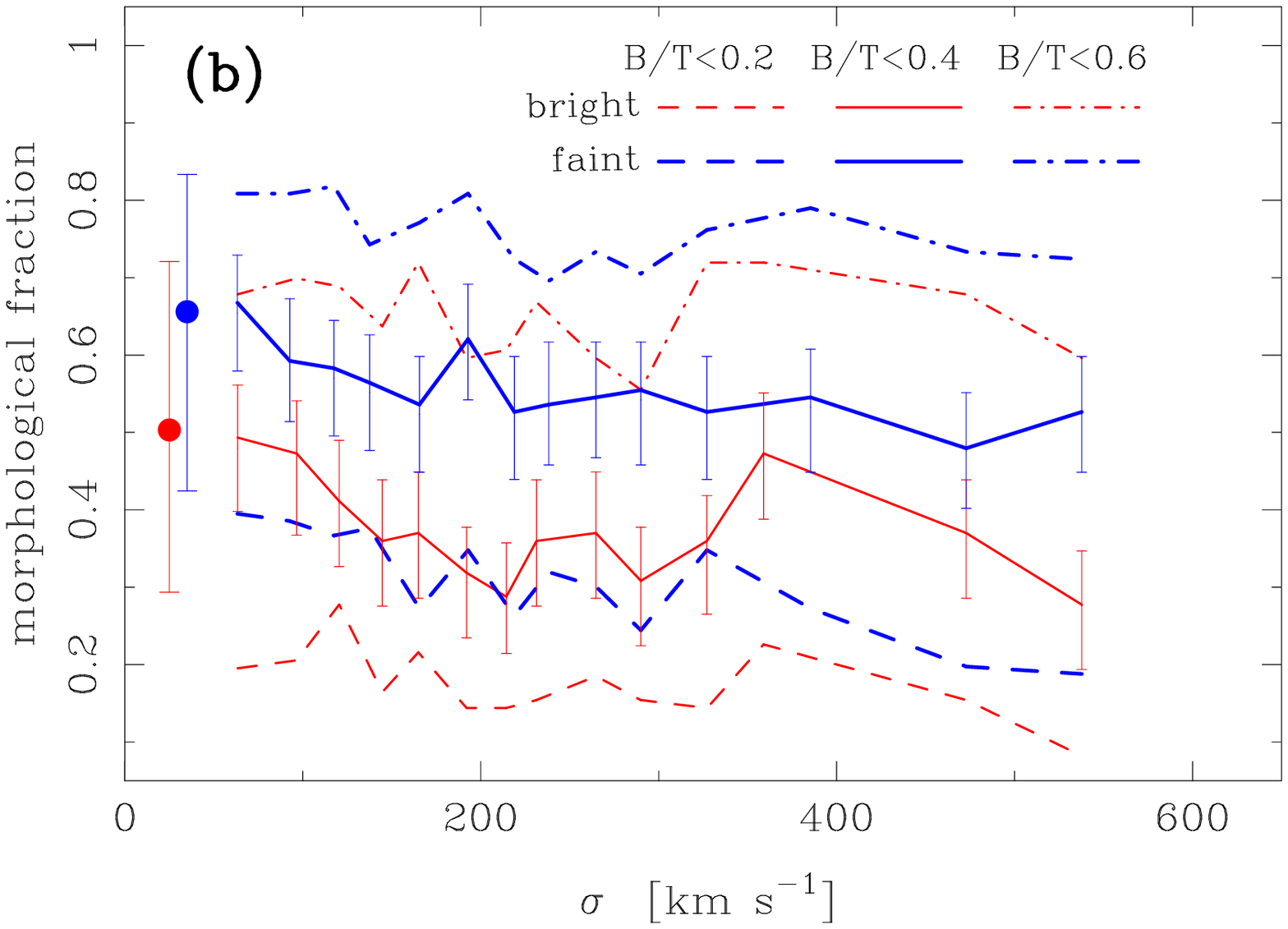}
\caption{
{\bf (a)} EW(H$\alpha$) and {\bf (b)} morphological fractions plotted
as a function of velocity dispersion.
The red and blue lines mean bright and faint galaxies, respectively.
Each bin contains approximately 100 galaxies.
In (a), the lines show the median and the quartiles of the distribution.
In (b), the lines show the morphological fractions of
$B/T<0.2,\ 0.4,\ \rm and \ 0.6$, respectively.
The leftmost points and the associated error bars in each panel indicate
the median and quartiles of the distribution of field galaxies (i.e., galaxies outside of the FOFA groups).
The other error bars show the bootstrap $90\%$ intervals.
\label{fig:sigma_dep}
}
\end{figure}

We examine dependencies of galaxy properties on velocity dispersion.
As shown in Figure \ref{fig:sigma_dep},
the star formation and morphology of bright galaxies do not significantly change with system richness,
and bright galaxies are mostly non--star-forming early-type galaxies.
As for faint galaxies, albeit with the large errors,
we cannot find a clear trend in morphology.
But it seems that galaxies become morphologically earlier in richer systems.
Note that faint galaxies are morphologically later than bright galaxies in all systems.
Most of the faint galaxies in systems with $\sigma\gtrsim200\kms$
are non-star-forming galaxies (see the median line in the figure).
Galaxies in systems as poor as $\sigma\sim100\kms$ seem to show weaker star formation activities
than field galaxies.
To clarify this, we compare the \ewha distribution of galaxies in $\sigma=75-125\kms$ systems
with that of field galaxies.
The K-S probabilities are 0.4\% for bright galaxies and 0.02\% for faint galaxies.
Therefore, we emphasize that galaxy star formation in systems as poor as $\sigma\sim 100\kms$
is less active compared with the field.
The fraction of star-forming galaxies becomes slightly larger
in poor systems ($\sigma\lesssim200\kms$).
However, such poor systems may contain a non negligible fraction of chance interlopers,
and the small change may not be real (we do not quantify the amount of the contamination here).
One may concern an effect of an error in our velocity dispersion estimates on our results.
We find that our results are essentially unchanged if we use the FOFA systems with
a relatively small bootstrap error in velocity dispersion ($\Delta\sigma/\sigma<0.2$).

\begin{figure}
\plotone{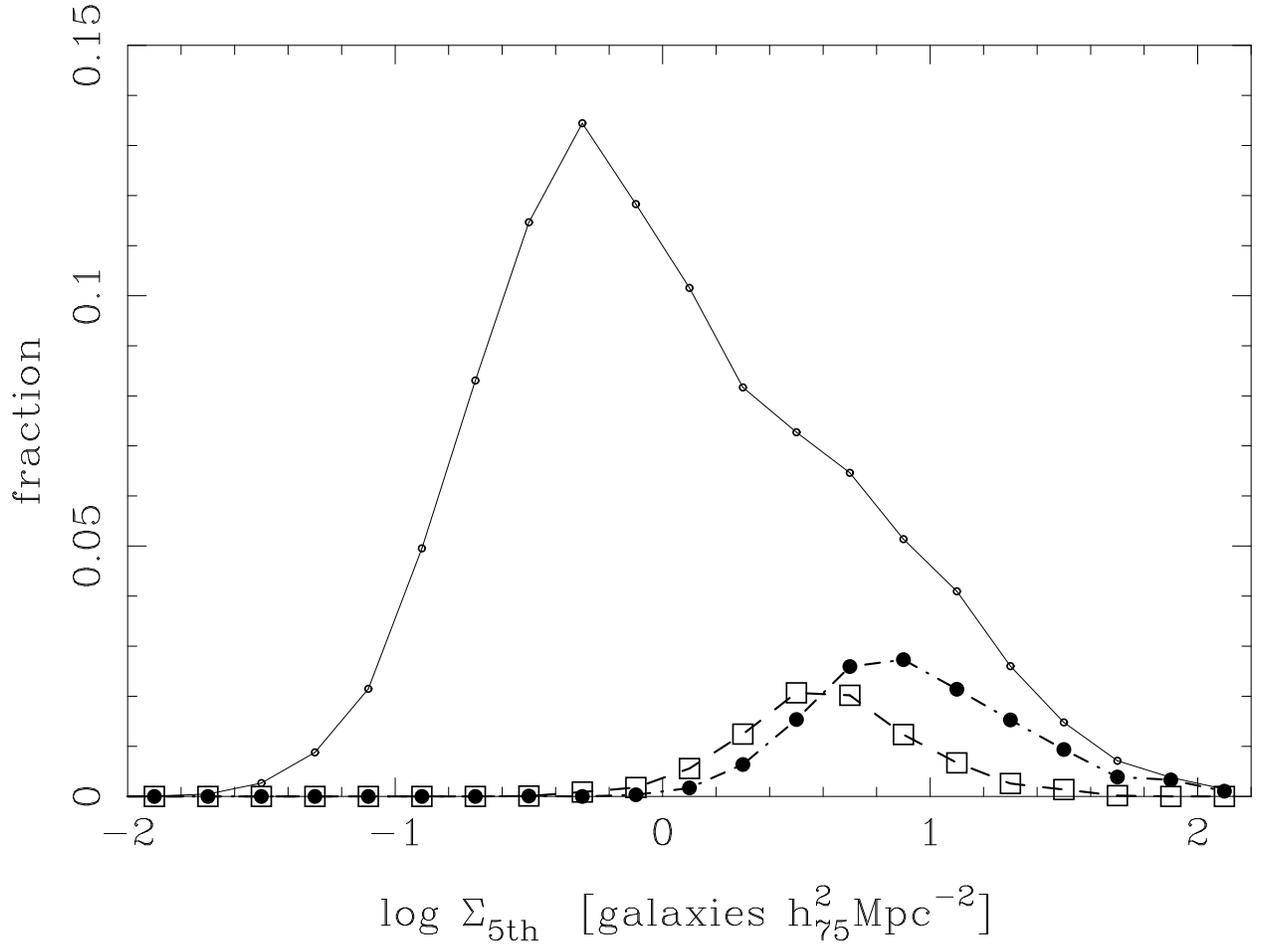}
\caption{
Local density distribution of the entire sample (solid line),
the FOFA systems with $\sigma<200\kms$ (dashed line),
and the FOFA systems with $\sigma>200\kms$ (dot-dashed line).
\label{fig:local_sigma}
}
\end{figure}

To clarify the relationship between this analysis and the discussion presented in the
previous section, we show in Figure \ref{fig:local_sigma} the local density distribution
of the entire sample, the FOFA systems with $\sigma<200\kms$, and the FOFA systems
with $\sigma>200\kms$.
It can be seen that galaxies in the FOFA systems have high local density,
and a significant fraction of galaxies at $\Sigma_{\rm 5th}>\Sigma_{\rm crit}$
belongs to the FOFA systems.
This means that the break in properties of faint galaxies presented in the previous section
is driven by the galaxy systems, and therefore the critical environment where faint galaxies change
their properties is galaxy systems.
Galaxies in rich systems tend to have higher local density than those in poor systems.
This might reflect the fact that galaxies in systems with $\sigma<200\kms$ show more active
star formation than those in systems with $\sigma>200\kms$.

\subsection{Summary of the Dependencies on System Richness}

We run the FOFA to find galaxy systems and measure their velocity dispersions.
Bright galaxies show little dependence on system richness,
and they are mostly non-star-forming early-type galaxies.
Faint galaxies seem to show a weak change in their properties in very poor systems ($\sigma < 200\kms$),
although the change may be driven, at least in part, by the contamination of chance interlopers and
unphysical associations.
An important point is that galaxies in $\sigma \gtrsim 200\kms$ systems are mostly non-star-forming galaxies,
and even galaxies in $\sigma\sim100\kms$ systems show weaker star formation activities
than those in the field.
We find that the FOFA systems dominate the local density region of $\Sigma_{\rm 5th}>\Sigma_{\rm crit}$.
The break in properties of faint galaxies reported in the previous section
is driven by the circumstance that galaxies in systems have
weaker star formation activities and earlier morphological types than those in the field.

Our results can be summarized as follows (see also Table 3).
\begin{enumerate}
\item The star formation and morphology of bright galaxies show nearly no correlation with system richness.
\item Faint galaxies are dominated by non-star-forming galaxies at $\sigma > 200\kms$.
\item Even in systems with $\sigma\sim100\kms$, galaxies show weaker star formation activities
than those of field galaxies.
\item The star formation and morphology of faint galaxies show small changes
at $\sigma < 200\kms$, but this may be caused by contamination by chance interlopers.
\item Faint galaxies are morphologically later than bright galaxies in all systems.
\item Galaxy systems dominate the local density range of $\Sigma_{\rm 5th}>\Sigma_{\rm crit}$.
Thus, galaxy systems should be the driver of the break in the properties of faint galaxies
reported in the previous section.
\end{enumerate}

\section{DISCUSSION}


\subsection{Comparison with Previous Studies}

\citet{lewis02}, \citet{gomez03}, and \citet{balogh04} reported
that star formation in galaxies drops at $\Sigma_{\rm local}\sim 1 \ \rm galaxies{\it\ h}_{75}^{2}\rm\ Mpc^{-2}$,
although the definition of local density differs from author to author.
We have identified the critical density of $\Sigma_{\rm 5th}\sim 2.5\ \rm galaxies{\it\ h}_{75}^{2}\rm\ Mpc^{-2}$
($\log \Sigma_{\rm 5th}\sim 0.4$).
The difference in the critical density between our estimate and the previous estimates
is due to the difference in the magnitude cut in the sample construction.
\citet{lewis02} used galaxies brighter than $M^*+0.75$, and
\citet{gomez03} and \citet{balogh04} adopted a magnitude cut of $M^*+1$.
Thus the previous estimates are based on samples shallower by $\sim1$ mag than our sample.
On the basis of the luminosity function given by \citet{blanton01},
the galaxy density in our sample is estimated to be
twice as high as those in the previous authors.
Therefore, our critical density appears to be in good agreement with the previous estimates.

However, the previous found a break for $M_r\lesssim M_r^*+1$ galaxies
(i.e., bright galaxies in our definition) for which we have identified no clear break
(e.g., see Fig. \ref{fig:local_vs_ha}).
Our result is that only faint galaxies show a break.
This discrepancy is possibly due to the quantitative difference in the definition of local density
(but note that Balogh et al. 2004 adopted the same definition as ours).
It should be noted that the previously reported breaks are not as prominent as the one we have found.
This is probably because fainter galaxies tend to show a stronger break.

\citet{martinez02} claimed that there is a strong correlation between the relative fraction of star forming
galaxies and the virial mass of the parent system.
Since their spectral classification is based on the PCA \citep{madgwick02} and
thus their definition of star-forming galaxies is different from ours,
their results cannot be directly compared with our results.
It is known, however, that there is a clear correlation between their spectral index ($\eta$) and EW(H$\alpha$).
Adopting their threshold $\eta=-1.4$ as EW(H$\alpha)=0\rm \AA$ \citep{madgwick02},
we find that the correlation mostly arises from poor systems,
which are subject to interlopers and spurious systems.
If the inner two bins in their Figure 2 (which roughly correspond to $\sigma\lesssim200\kms$) are excluded,
the remaining correlation is not very significant ($<2\sigma$).
It should be noted that their sample reaches $\sim M^*+2.5$, and the galaxies are selected by $b_J$ magnitude.
Therefore, their sample, as compared with ours, is biased toward star--forming galaxies.
They also concluded that star formation in group galaxies is strongly suppressed compared with field galaxies
even for very poor groups with $M_{virial}\sim10^{13} M_\odot$,
which corresponds roughly to $\sigma\sim250\kms$.
This finding is in good agreement with our result.

\citet{kauffmann04} reported that, at a fixed stellar mass,
star formation and nuclear activity depend strongly on
local density, while morphology is almost independent of local density.
But low--mass galaxies ($M_{stellar}<3\times10^{10}M_\odot$) show morphological change
in the highest--density environment.
Kauffmann et al.'s  study is based on stellar masses of galaxies, while ours is based on
luminosities of galaxies.
Thus, direct comparisons cannot be made.
It is, however, worthwhile to investigate the correspondence between our magnitude cut
for separating bright and faint galaxies ($M_r=M_r^*+1$) and the stellar mass.
Using the prescription of \citet{baldry04}, we find that our magnitude cut corresponds to
$\log (M_{stellar}/M_\odot)\sim10.6$ for red galaxies and $\log (M_{stellar}/M_\odot)\sim10.2$ for blue galaxies.
\citet{baldry04} found that red galaxies dominate the stellar mass range of
$\log (M_{stellar}/M_\odot)>10.2$, while blue galaxies dominate the lower stellar mass range.
This should reflect the fact that
bright galaxies are redder than faint galaxies in a statistical sense (Fig. 2).
We find that morphology changes as a function of local density, even for bright galaxies.
It is not immediately clear whether this finding is inconsistent with \citet{kauffmann04}
since they use the concentration index, which is subject to the seeing effect, as shown in \S 3.3,
and their definition of local density is different from ours.
We do not try to compare the results here.
It is found in \S 3 that faint galaxies show the break at the critical density, while bright galaxies do not.
This may be related to the fact that our magnitude cut for separating bright and faint galaxies corresponds to
$\log (M_{stellar}/M_\odot)\sim10.2$ (for blue galaxies), where the midpoint of the transition
of galaxy properties is reported to occur \citep{baldry04}.

\subsection{Physical Interpretations}
\subsubsection{Proposed Mechanisms}
It has long been known that galaxies in rich systems follow a different evolutionary path
from those in the field.
There are many non--star-forming early-type galaxies in rich systems,
whereas star-forming spiral galaxies are generally found in the field
\citep{dressler80,balogh97,hashimoto98,balogh99,lewis02,gomez03,goto03c,balogh04}.

Several mechanisms have been proposed to drive the observed environmental dependence of galaxy properties.
Ram pressure stripping is one of the environmental mechanisms expected to be effective
in the cores of rich systems \citep{gunn72,abadi99,quilis00}.
When a galaxy moves through the hot intracluster medium at a high speed,
it feels a ``wind,'' and the cold gas in it will be stripped off.
Moreover, galaxies in rich systems experience high--velocity encounters with other galaxies (harassment).
Should such encounters occur successively, galaxy disks would be destroyed and
starbursts would be triggered \citep{moore96a,fujita98,moore99,mihos03}.
Interactions with the gravitational potential of the system should also be significant \citep{byrd90}

In semi-analytic galaxy evolution models, galaxies are assumed to have hot halo gas,
and the gas is lost when they sink into a larger halo like a group or a cluster (e.g., \citealt{okamoto03}).
If the halo gas is the source of the cold gas in the disk, then
the lack of halo gas leads to a slow decline in SFR ($>1\rm Gyr$)
as the galaxy consumes the remaining cold gas \citep{larson80,balogh00,diaferio01}.
This mechanism is often referred to as strangulation, suffocation, starvation, or halo--gas stripping.
Here we call it strangulation.
Strangulation models have achieved great success in reproducing star formation in galaxies
in rich systems \citep{balogh00,diaferio01,okamoto03}.
A remarkable feature is that the radial distribution of star formation strength
in rich systems can be reproduced by the strangulation effect.

Low-velocity interactions between galaxies will trigger starbursts consuming the cold gas
in the galaxies and weaken subsequent star formation activity.
Actually, galaxies in close pairs show enhanced star formation rates over those of isolated galaxies,
and probably they are currently in the process of interaction \citep{patton97,barton00,lambas02}.
The morphology of galaxies will be more or less disturbed depending on the strength of an interaction.
For an extreme case, a major merger will produce a single elliptical galaxy
after strong starbursts (e.g., \citealt{mihos96}).

The above mechanisms can be categorized into two broad classes:
mechanisms that are effective only in rich systems (ram pressure stripping of cold gas and harassment),
and mechanisms that are effective in other environments, such as poor systems and the field
(strangulation and low-velocity interactions).
In the following subsections,
we discuss implications of our findings in regard to putting constraints on the proposed mechanisms and galaxy evolution.
First, we focus on the dependence of galaxy properties on system richness because the
effects of the proposed mechanisms are expected to be a strong function of system richness.
Then, implications of the dependencies on local density are discussed to put further constraints.
Finally, we briefly address galaxy evolution.

\subsubsection{System Richness Dependencies}
In \S 4, we showed that non--star-forming galaxies make up the dominant
population even in galaxy systems as poor as $\sigma\sim200\kms$ (Table 3).
Now we ask what mechanism can explain this observation.
Numerical studies have shown that ram pressure stripping of cold gas
is only effective in the cores of rich systems where galaxies are moving at high speeds and
the hot intracluster medium is abundant
\citep{abadi99}.
The same is true for harassment \citep{moore96b}.
These mechanisms cannot be effective in systems as poor as $\sigma\sim200\kms$.
We recall that galaxies in systems as poor as $\sigma\sim100\kms$ show
weaker star formation activities than those in the field.
Accordingly, the dominant population in such poor systems cannot be fully explained by those mechanisms.
Therefore, it is likely that any mechanisms that are effective only in rich systems
have not played a major role in transforming galaxies into non--star-forming galaxies.
Rather, mechanisms that are effective in the field and/or in poor systems,
such as low-velocity interactions and strangulation, are the candidates of interest.

\subsubsection{Local Density Dependencies}

To obtain a deeper insight into the mechanisms involved, we focus on passive spirals
\citep{couch98,dressler99,poggianti99,goto03d} or, equivalently, anemic spirals \citep{vandenbergh76}.
Passive spirals are considered to be products of ram pressure stripping of cold gas or strangulation,
since these mechanisms suppress star formation activity in galaxies
but do not strongly disturb their morphology.
The reader is referred to \citet{goto03d} for an extensive analysis of passive spirals.

\begin{figure}
\plottwo{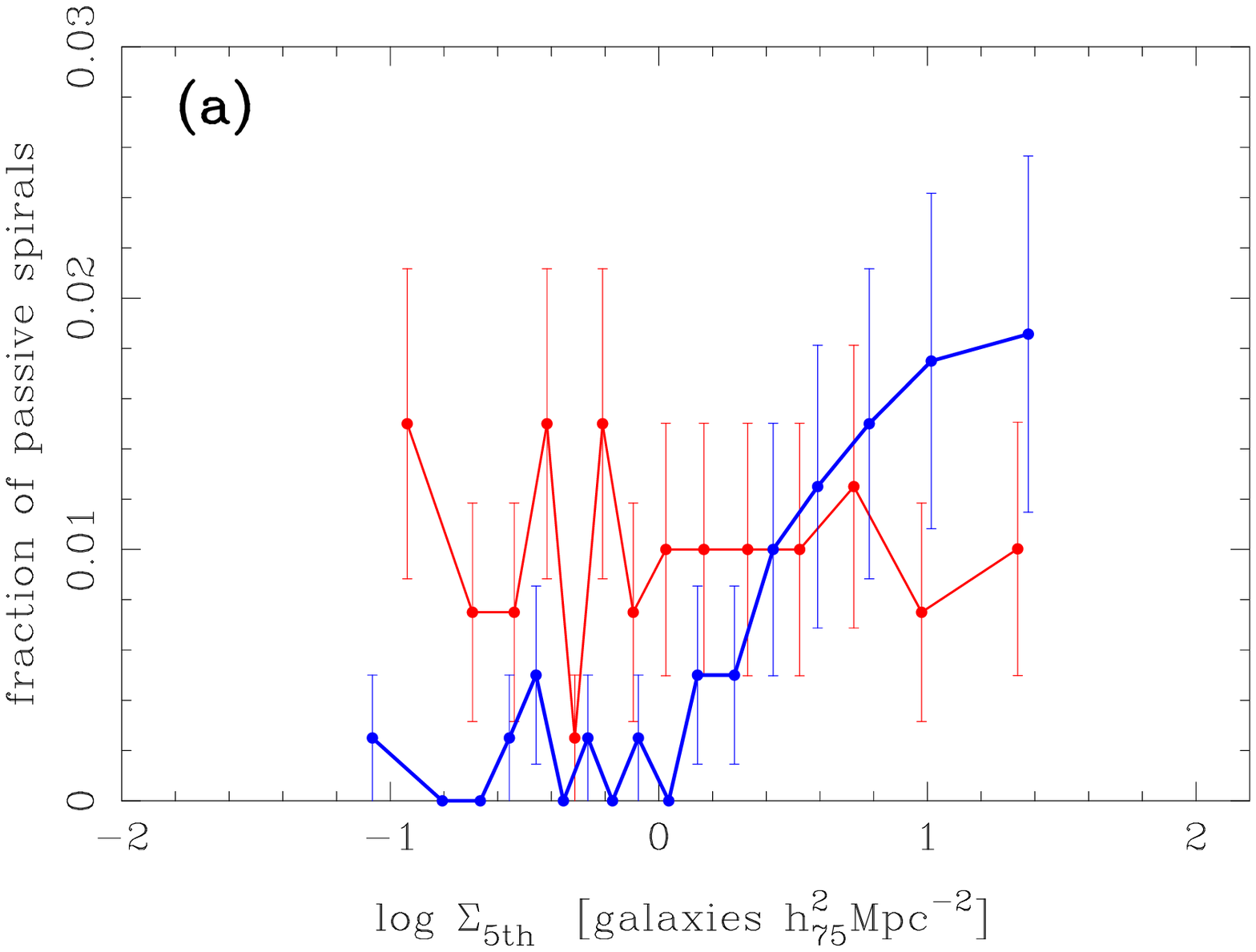}{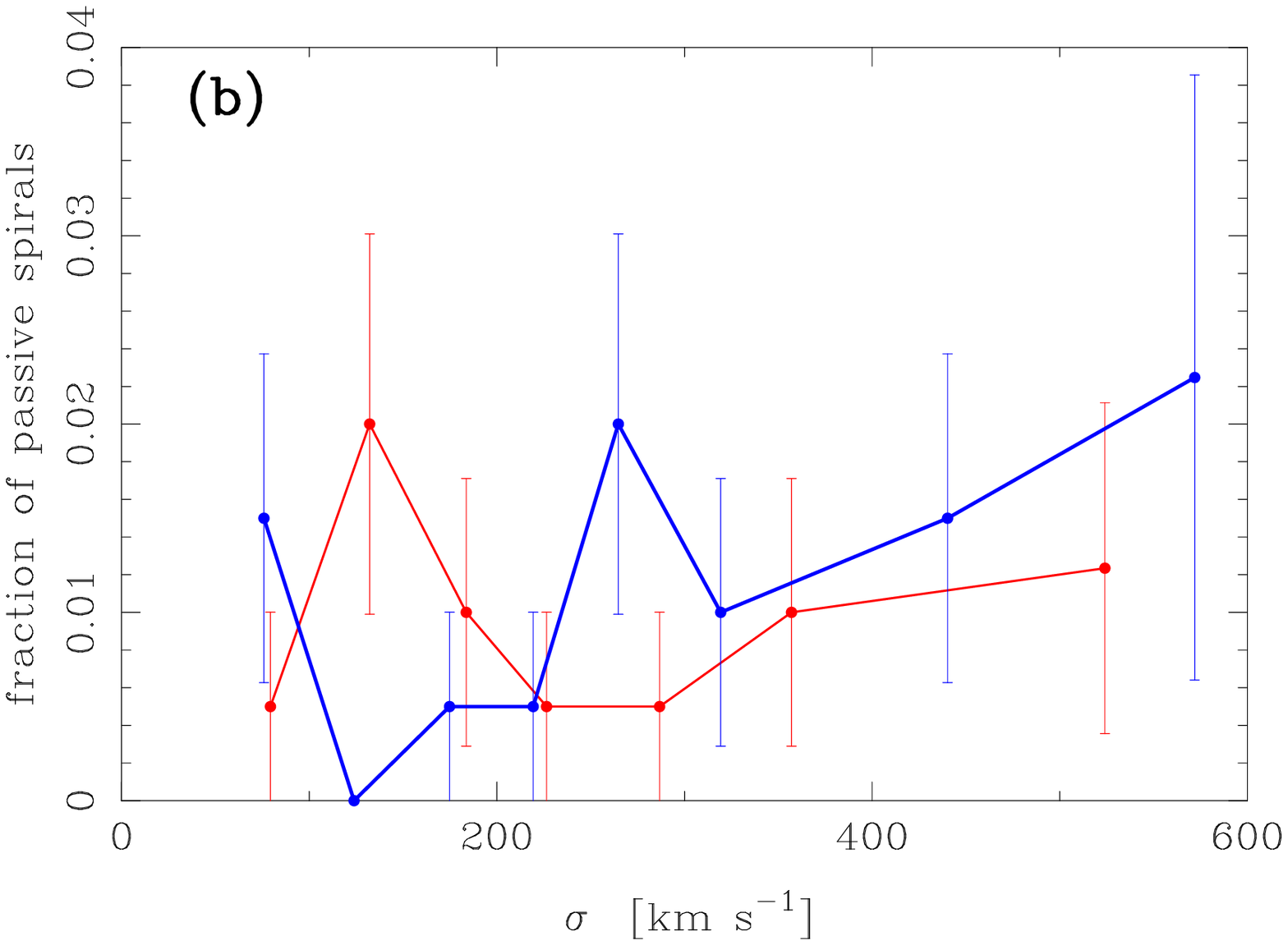}
\caption{
{\bf(a)}
Fraction of passive spirals as a function of local density.
Passive spirals are defined as galaxies with $\rm EW(H\alpha)<0$ and $B/T<0.2$.
The red and blue lines represent bright and faint galaxies, respectively.
The error bars show the $1\sigma$ errors based on the Poisson statistics.
We note that the relative fraction of bright and faint passive spirals should not be
directly compared because of the strong aperture bias.
{\bf(b)} Fraction of passive spirals as a function of velocity dispersion of galaxy systems.
\label{fig:passive_spirals}
}
\end{figure}

We define passive spirals as those with $\rm EW(H\alpha)<0$ and $B/T<0.2$.
The population fraction of passive spirals as a function of local density
is shown in Figure \ref{fig:passive_spirals}(a).
Bright passive spirals live in all local environment uniformly.
On the other hand, faint passive spirals preferentially reside in dense regions,
especially regions denser than the critical density ($\log\Sigma_{\rm 5th}>0.4$).
It is therefore suggested that ram pressure stripping of cold gas and/or strangulation
is at work on faint galaxies in dense environment.

To discriminate between ram pressure stripping of cold gas and strangulation,
the population fraction of passive spirals as a function of system richness is examined.
As shown in Figure \ref{fig:passive_spirals}(b),
the fraction of faint passive spirals seem to increase with system richness.
However, the trend should not be over interpreted because of the poor statistics.
The K-S test shows that the distribution of passive spirals is not very different
from that of the parent sample (the K-S probabilities of 80\% for bright galaxies
and 23\% for faint galaxies).
The interesting point here is that we find passive spirals in poor systems ($\sigma\lesssim300\kms$),
as well as in richer systems.
Passive spirals in such poor systems are not expected to be products
of the ram pressure stripping of cold gas.
Therefore, it is likely that strangulation actually plays a role
in suppressing star formation in faint galaxies.
However, if so, it is puzzling that bright galaxies are not affected by strangulation.

Since properties of bright galaxies have no break at the critical density
and their properties show a monotonic change
over the entire local density range, their environmental dependencies may be determined by
a mechanism that works even in the field, namely, low-velocity interactions.
The fact that morphology of bright galaxies strongly changes with local density supports this view (Figure 5).
We have shown that the brightest galaxies ($M_r<M_r^*-1$) are all red early-type galaxies,
independent of environment.
This would be additional supporting evidence because the final products of interactions and mergers
are expected to be luminous early-type galaxies.
However, some points are left unexplained.
For example, if the morphology-density relation of bright galaxies is a product of low-velocity interactions,
why does the morphology of faint galaxies fail to show any change below the critical density?
One possible interpretation is that different mechanisms affect galaxies of different luminosities.
That is, the primary effect on bright galaxies is low-velocity interactions,
while that on faint galaxies it is strangulation.
There is, however, no theoretical background for this interpretation.
Further studies are required to pursue the issue further.

\subsubsection{Environmental Effects in the Local universe}
If galaxies are currently under the influence of environmental effects
that cause the slow transformation of galaxy properties (e.g., strangulation),
we expect to see a signature of the transformation
(e.g., reduced star formation rates in star-forming galaxies).
In \S 3, we showed that the distribution of the star formation rates of
bright star-forming galaxies [EW(H$\alpha)>4\rm\AA$] does not change with local density.
Similarly, morphology of bright late-type galaxies ($C_{in} > 0.45$) does not change with local density.

There are at least three ways to interpret this result.
One is that the time scale of the transformation of bright galaxies is too short
to observe the transition phase.
Another is that the transformation of bright galaxies occurs in all environments uniformly.
The other is that the transformation of bright galaxies does not frequently occur in the local universe.
Unfortunately, we cannot clearly discriminate these possibilities quantitatively.
But, it can be said that there is no clear signature of the slow transformation of bright galaxies
in dense regions.

As for faint galaxies, star-forming galaxies show the reduced star formation rates above the critical density.
Morphology also seems to change.
This is a clear signature of the slow transformation in galaxy systems such as groups and clusters.
The fact that there are faint passive spirals in a dense environment offers supporting evidence.
But we do not mean that there is no rapid transformation;
the rapid transformation does not produce, for example, the reduced star formation rates in star-forming
galaxies, and we cannot put any constraint on it.
To summarize, we suggest that the slow transformation of faint galaxies occurs to some extent
in a dense environment in the local universe.
The transformation of bright galaxies is poorly constrained with the results at hand, but
there is no clear signature of the slow transformation of bright galaxies in dense environment.

\subsubsection{Implications for Galaxy Evolution}
Because galaxies in the local universe are ``final'' products of galaxy evolution over the Hubble time,
it is interesting to discuss what our findings tell about galaxy evolution.
As shown in \S 4.2, the environment denser than the critical density corresponds to galaxy systems
(see also Appendix E).
Thus, the presence of a break at this critical density means that galaxies are affected by galaxy systems.
If galaxies are affected by system-specific mechanisms,
red early-type galaxies will be accumulated in galaxy systems over time.
Accordingly, we expect to observe a strong change in the galaxy population around the critical density.
This is what we see in faint galaxies (Fig. \ref{fig:local_vs_ha4}).
The fact that bright galaxies show no break at the critical density suggests that their evolution is
not strongly related to galaxy systems such as groups and clusters.
Therefore, it seems that bright and faint galaxies have followed different evolutionary paths.
It is, however, unclear why the influence from systems weakens with increasing galaxy luminosity.
We need to study fainter galaxies and galaxies at higher redshift to address this point.
Comparisons with semi-analytic models that take into account the effects of
both interactions and strangulation (e.g., Okamoto \& Nagashima 2003) will improve our understanding.
We leave them for future work.

\section{CONCLUSIONS}


We have examined the environmental dependence of star formation and morphology of galaxies
on the basis of data from the SDSS.
Galaxies are restricted to the redshift range of $0.030<z<0.065$ and
the magnitude range of $M_r<M_r^*+2\ (=-19.4)$, which is deeper by 1 mag
than previous volume-limited studies \citep{lewis02,gomez03,goto03c,balogh04}.
We adopt $g-i$ color and \ewha as star formation indicators,
as well as \bt \citep{okamura99} and \cin \citep{shimasaku01}, as morphology.
Our sample is divided into two subclasses by the absolute magnitude of $M_r^*+1$,
as shown in Table 1.

We investigated the dependence of galaxy properties on local density.
There is a critical density ($\log\Sigma_{\rm 5th}\sim0.4$)
at which both star formation and morphology abruptly change.
The break at the critical density is seen only for faint galaxies, and
bright galaxies show no clear break (Figs. \ref{fig:local_vs_color}-\ref{fig:local_vs_bt}),
which appears to be inconsistent with previous studies \citep{lewis02,gomez03,goto03c,balogh04}.
We have thus found that galaxies of different luminosities have different dependencies on local density.
As an extreme case, the properties of the brightest galaxies ($M_r<M_r^*-1$)
show no correlation with local density at all.
All the correlations with local density are summarized in \S 3.6 and Table 2.

We also have focused on richness of galaxy systems.
We apply the friends-of-friends algorithm to find galaxy systems in our sample.
The statistical properties of galaxies have only weak dependence on system richness.
The only exception is a clear increase of star-forming late-type galaxies in very poor systems 
($\sigma<200\kms$; Figure \ref{fig:sigma_dep}), although
this increase is possibly caused by the contamination of chance interlopers.
We find that the FOFA systems dominate the local density region of $\Sigma_{\rm 5th}>\Sigma_{\rm crit}$,
and this should be the driver of the break.
The trends found as a function of richness are summarized in \S 4.3 and Table 3.

We suggest that mechanism that are effective in the field and/or in poor systems,
such as low-velocity interactions (e.g., mergers) and strangulation,
are the preferred candidates to explain the observed trends.
From the environmental distribution of passive spirals,
it is suggested that strangulation actually works on faint galaxies.
However, it is puzzling that bright passive spirals have no preference of environment.
The slow transformation (e.g., strangulation) of faint galaxies occurs to some extent
in a dense environment in the local universe,
but that of bright galaxies is not clearly seen in dense environment.
The fact that faint galaxies show a break in their properties at the critical density suggests
that their evolution is closely related to galaxy systems.
On the other hand, bright galaxies show no break and their evolution is expected to
have little connection with galaxy systems.
However, some unanswered questions are left and further investigations are required.

\section*{ACKNOWLEDGEMENTS}


We thank Tadayuki Kodama for a careful reading of an early version of the manuscript
and for useful comments that helped improve the paper.
M.T. would like to acknowledge Michael Balogh for useful discussion and
having the draft of his paper available.
We thank Masafumi Yagi and Nobuyuki Ohama for helpful comments.
We thank the referee, S. L. Morris, for his constructive comments which improved the paper.

Funding for the creation and distribution of the SDSS Archive has been
provided by the Alfred P. Sloan Foundation, the Participating Institutions,
the National Aeronautics and Space Administration, the National Science Foundation,
the U.S. Department of Energy, the Japanese Monbukagakusho, and the Max Planck Society.
The SDSS Web site is \url{http://www.sdss.org/}.
The SDSS is managed by the Astrophysical Research Consortium (ARC) for
the Participating Institutions. The Participating Institutions are The University
of Chicago, Fermilab, the Institute for Advanced Study, the Japan Participation Group,
The Johns Hopkins University, Los Alamos National Laboratory, the Max-Planck-Institute
for Astronomy (MPIA), the Max-Planck-Institute for Astrophysics (MPA), New Mexico State
University, University of Pittsburgh, Princeton University, the United States Naval
Observatory, and the University of Washington.





\appendix

\section{Fiber Collision}
Because of instrumental constraints, an SDSS spectroscopic fiber cannot be placed closer
than $55''$ to the neighboring fiber, resulting in reduced completeness in dense environments.
To correct for this incompleteness, we need to know the probability that a missed galaxy will
fall in the redshift range under consideration, and this is indeed very difficult to evaluate.
What we mean by a missed galaxy is a galaxy that is targeted
for the Main Galaxy Sample but not fed any fiber.

We can naively expect that the probability correlates, at least to some extent,
with the surface galaxy density in our redshift range.
If the position of a missed galaxy in the sky corresponds to field environment in our redshift range,
the probability will be low since galaxies are thinly populated around the missed galaxy.
On the other hand, if a missed galaxy falls in a cluster in our redshift range,
the probability will be high since the missed galaxy can be expected to be a cluster member.

We make a very rough estimate of the effect of the fiber collision on our density estimates.
It is somewhat difficult to evaluate the effect of the fiber collision in low-density environments
since the probability of the effect occurring is low and it is a non-trivial problem to know how low.
But the overall completeness of the SDSS spectroscopic survey is very high ($>90\%$), and
we expect that fiber collision is not a severe problem for field galaxies.
Thus, we focus on dense environments and evaluate the effect on our density estimates.
We select galaxies in environments denser than the critical density in our sample
and count missed galaxies that lie within $55''$ from the selected galaxies.
We find 1049 missed galaxies for 3505 selected galaxies.
Assuming all the missed galaxies fall in our redshift range, we find that we underestimate densities
above the critical density by a factor of $3505/(3505+1049)=0.77$ (i.e., approximately $-0.1$ dex) on average.
This should be considered a lower limit because not all the missed galaxies are in our redshift range.
The amount of the underestimation is relatively small, and therefore we conclude that
fiber collision does not significantly change our results.

\section{Aperture Bias}
Spectroscopic observation in the SDSS is performed through $3''$ diameter fibers, which are smaller than
a typical size of galaxies used in this paper.
Thus, a non negligible fraction of galaxy light is missed from each fiber, and
we inevitably underestimate the flux of galaxies.
This aperture bias is particularly significant when we analyze nebular emission lines, which are
expected to come mainly from a galaxy disk rather than a bulge.

\begin{figure}
\plotone{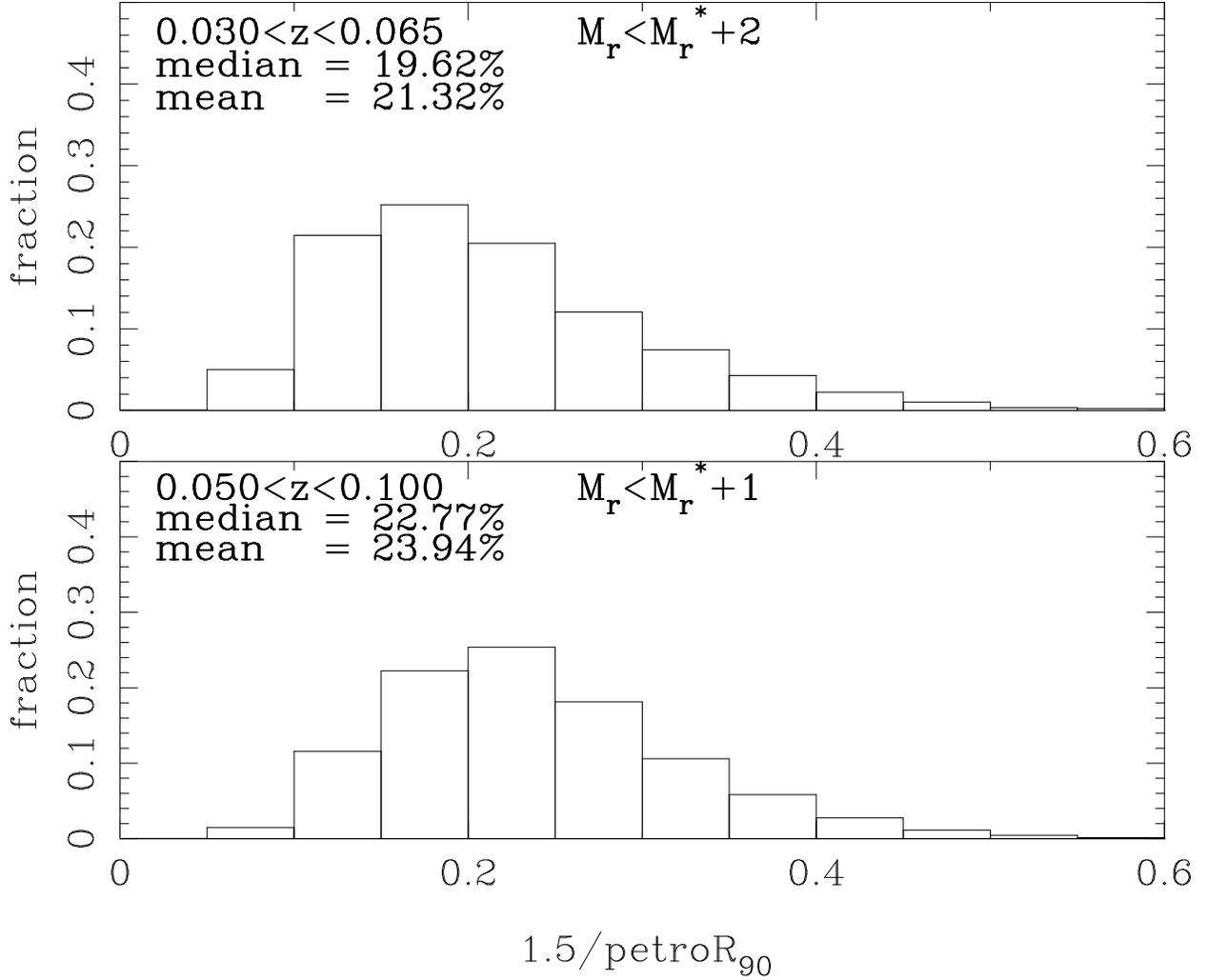}
\caption{
Distribution of the ratio of the fiber radius to the galaxy radius
for our sample (top panel) and a volume-limited sample of $0.050<z<0.100$ and $M_r<M_r^*+1$ (bottom panel).
\label{fig:fiber_gal_ratio}
}
\end{figure}

We present the distribution of the ratio of the fiber radius to the galaxy radius
in Figure \ref{fig:fiber_gal_ratio}.
We adopt Petrosian $90\%$ radius as a radius of a galaxy.
The ratio can be used for a very rough estimate of the amount of the aperture bias.
Of course, star-forming regions are not uniformly distributed within a galaxy.
Thus, the relative difference in the ratio does not simply reflect the relative strength of the aperture bias.
Note that  the environment might affect the distribution of star-forming regions in galaxies \citep{moss00}.
The top panel is for our volume-limited sample ($0.030<z<0.065$ and $M_r<M_r^*+2$), and
the bottom panel is for the volume-limited sample often used in the literature
($0.05<z<0.10$ and $M_r< M_r^*+1$).
In our sample, each fiber subtends only $\sim 20\%$ of the galaxy radius.
This means only $\sim 4\%$ of a surface area of a galaxy is included in a fiber.
Actually, each fiber will collect light from a larger area related to the effect of the seeing.
However, there is no doubt that we have a significant aperture bias in EW(H$\alpha$).
In the bottom panel, the ratio increases to $\sim 23\%$
(i.e., a $\sim 15\%$ increase compared with the top panel).
But even in this case, a fiber covers only $\sim 5\%$ of a surface area of a galaxy.

\begin{figure}
\plotone{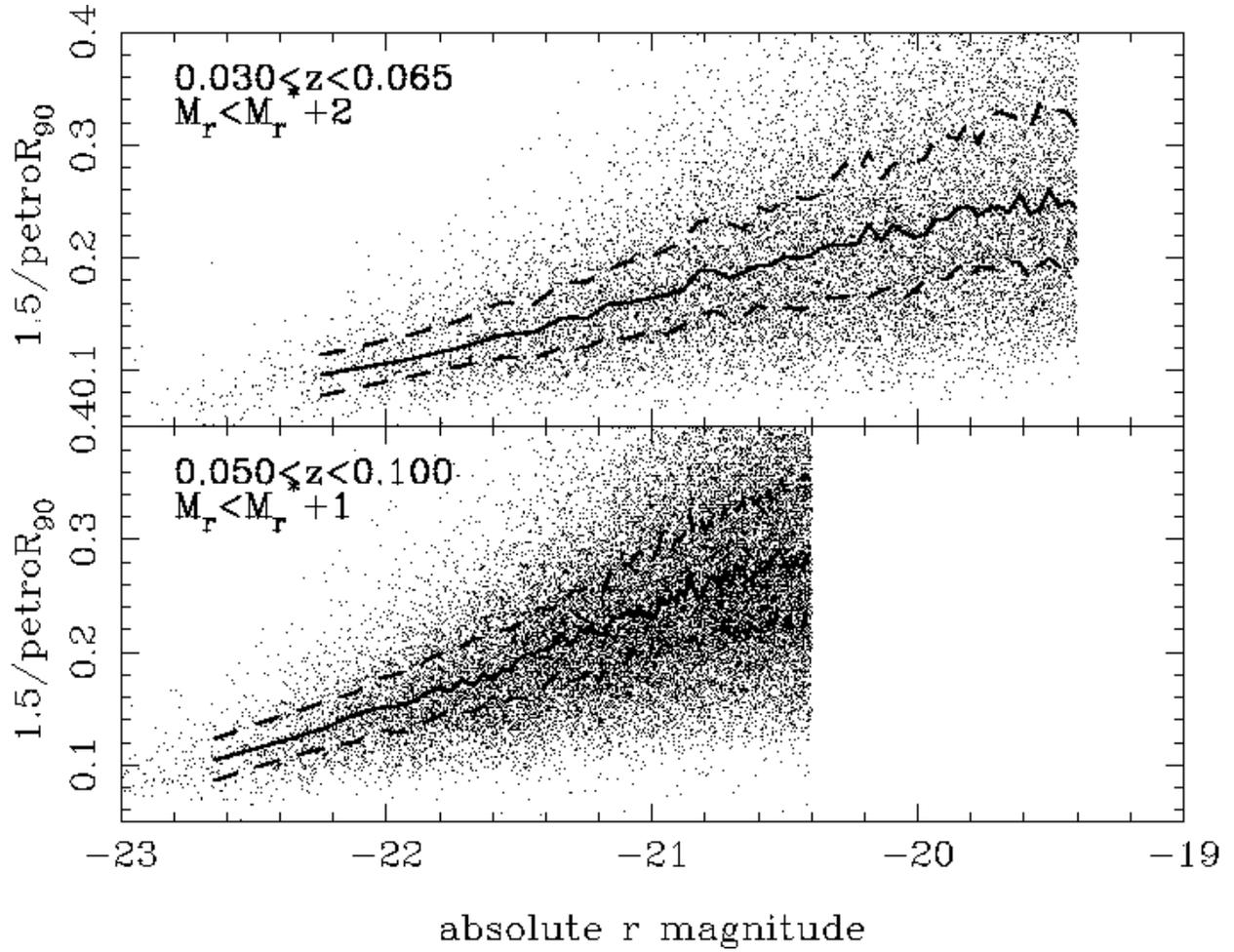}
\caption{
The fiber-galaxy radius ratio is plotted against absolute magnitude for our sample (top panel)
and the volume-limited sample of $0.050<z<0.100$ and $M_r<M_r^*+1$ (bottom panel).
The solid and dashed lines show the median and quartiles of the distribution.
\label{fig:mag_vs_ratio}
}
\end{figure}

Since the size of a galaxy is a function of luminosity, the aperture bias is a function of luminosity.
As shown in Figure \ref{fig:mag_vs_ratio}, the fiber--galaxy radius ratio changes with absolute magnitude
($\Delta ratio\gtrsim 15\%$ for our case).
We investigated the luminosity dependence of EW(H$\alpha$) in \S 3, but the results
should not be over interpreted because of the strong aperture bias.
However, the fact that color (which is free from the fixed-aperture bias) and EW(H$\alpha$) show almost
identical environmental dependencies suggests that most of the dependencies of EW(H$\alpha$) are not
driven by the aperture bias.
It is interesting to note that the aperture bias is a stronger function of luminosity 
in the volume-limited sample of $0.05<z<0.1$, on which most of the previous studies are based.

\section{Environmental Dependence of Absolute SFR}

We present the environmental dependence of absolute SFR.
There are several formulae in the literature to convert H$\alpha$ flux to SFRs
(e.g., \citealt{kennicutt98,hopkins01,charlot01,hopkins03}).
We follow the recipe of Hopkins et al. (2003, their eq. [B2]) to correct for
dust extinction and fixed-fiber aperture, and estimate absolute SFRs.
The reader is referred to that paper for details on the procedure.

\begin{figure}
\plottwo{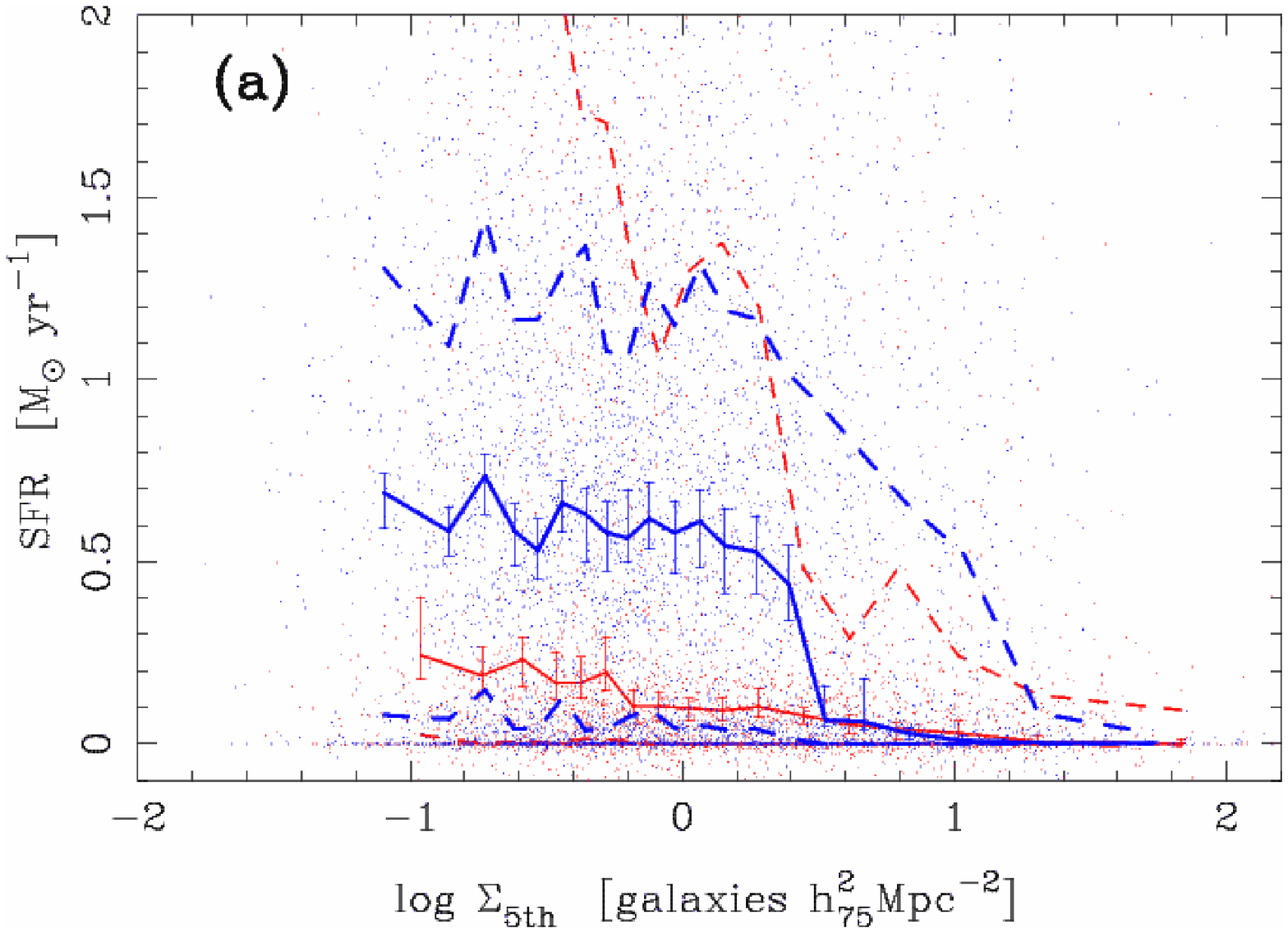}{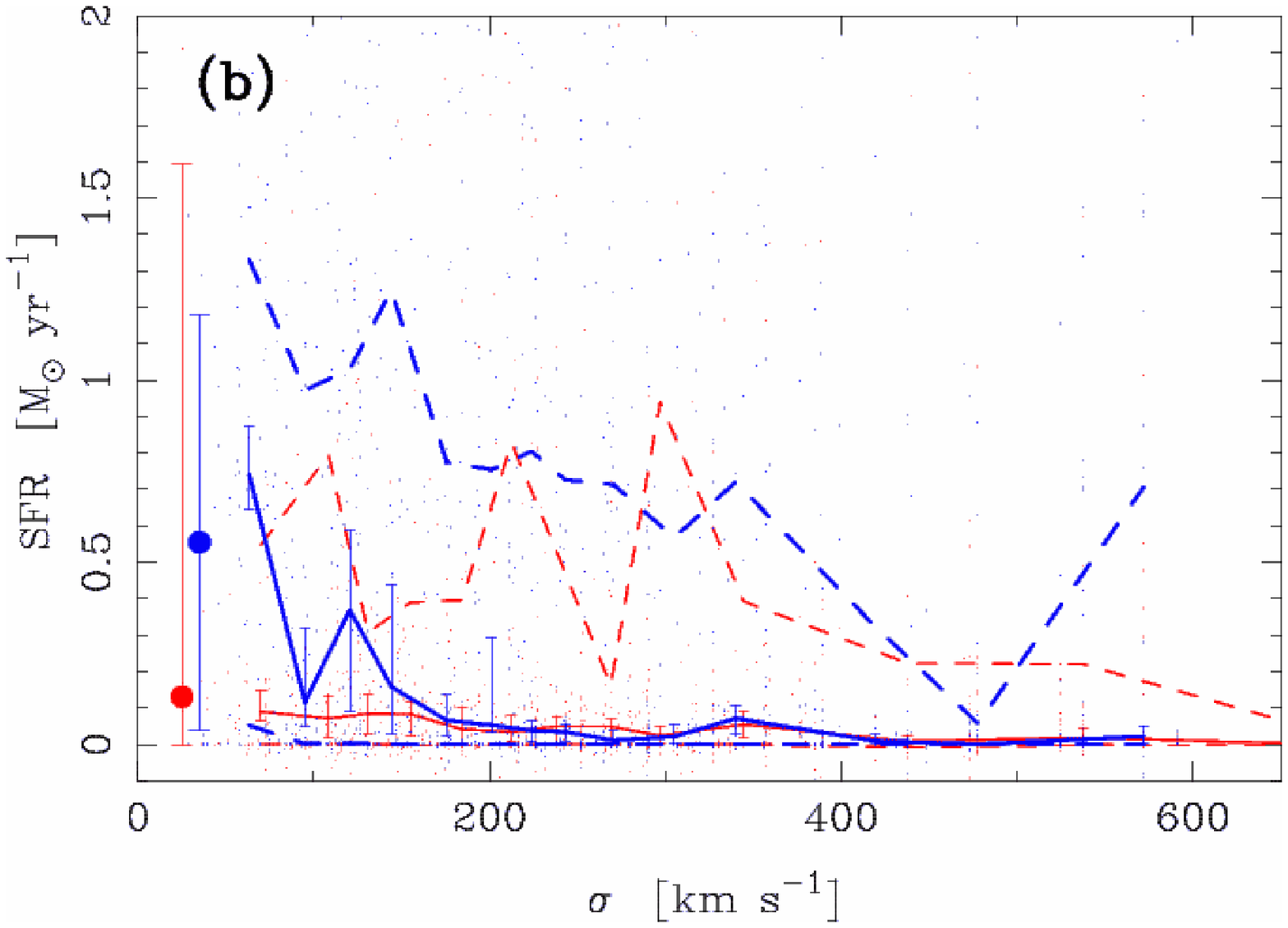}
\caption{
{\bf (a)} Same as Figure \ref{fig:local_vs_ha}, but for absolute SFR.
{\bf (b)} Same as Figure \ref{fig:sigma_dep}, but for absolute SFR.
Note that in both panels the error bars represent only statistical errors, and the errors in SFR estimates
are not included.
\label{fig:env_vs_sfr}
}
\end{figure}

The absolute SFR as a function of local density is presented in Figure \ref{fig:env_vs_sfr}(a).
The median SFRs show a similar trend to that seen in Figure 3; bright galaxies show no clear break whereas
faint galaxies have a strong break.
The dependence of the absolute SFR on system richness is shown in Figure \ref{fig:env_vs_sfr}(b).
The trend is essentially the same as that found in Figure \ref{fig:sigma_dep}.

\section{Dependence on Local Density and Luminosity}

\begin{figure}
\plottwo{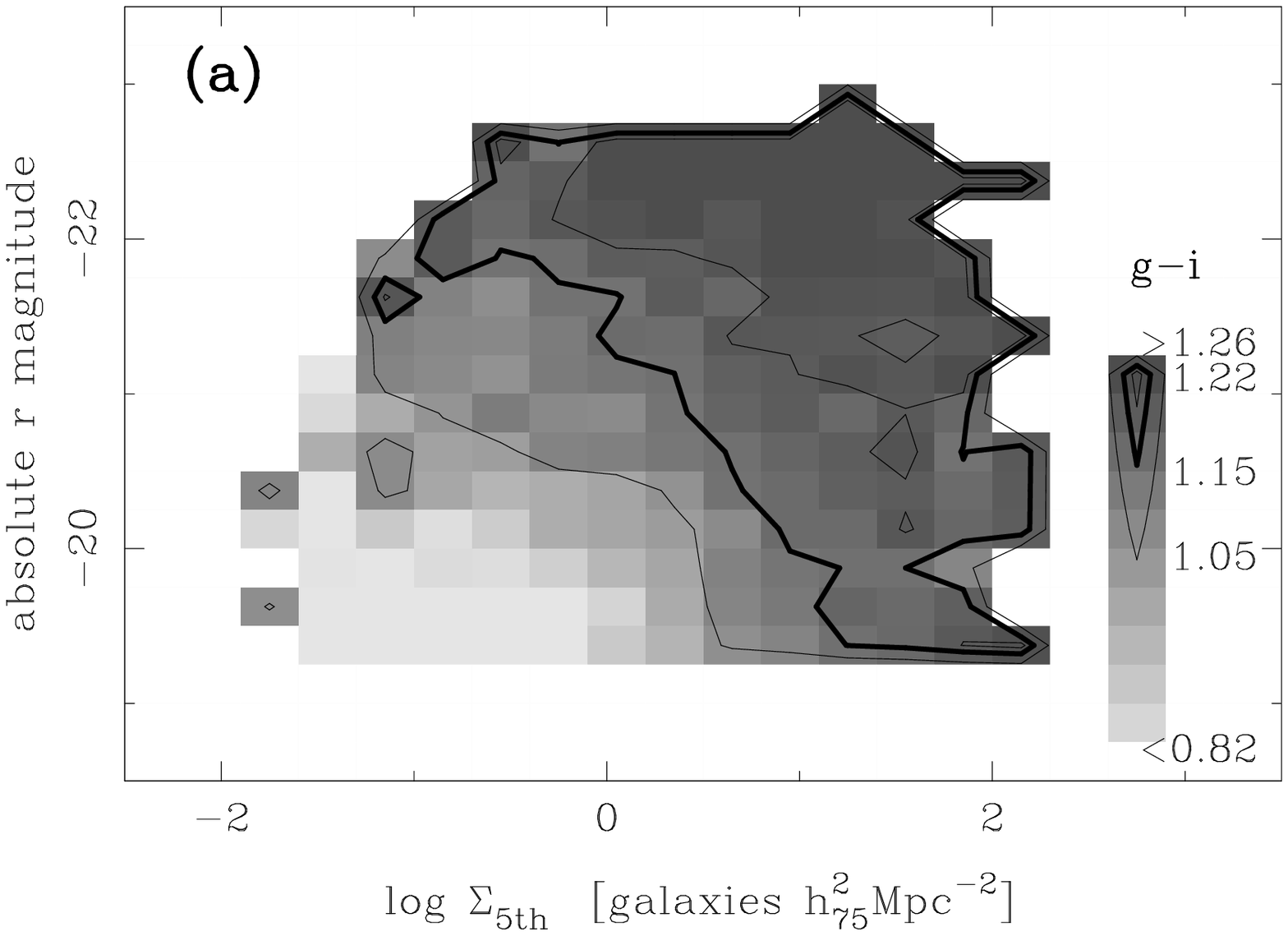}{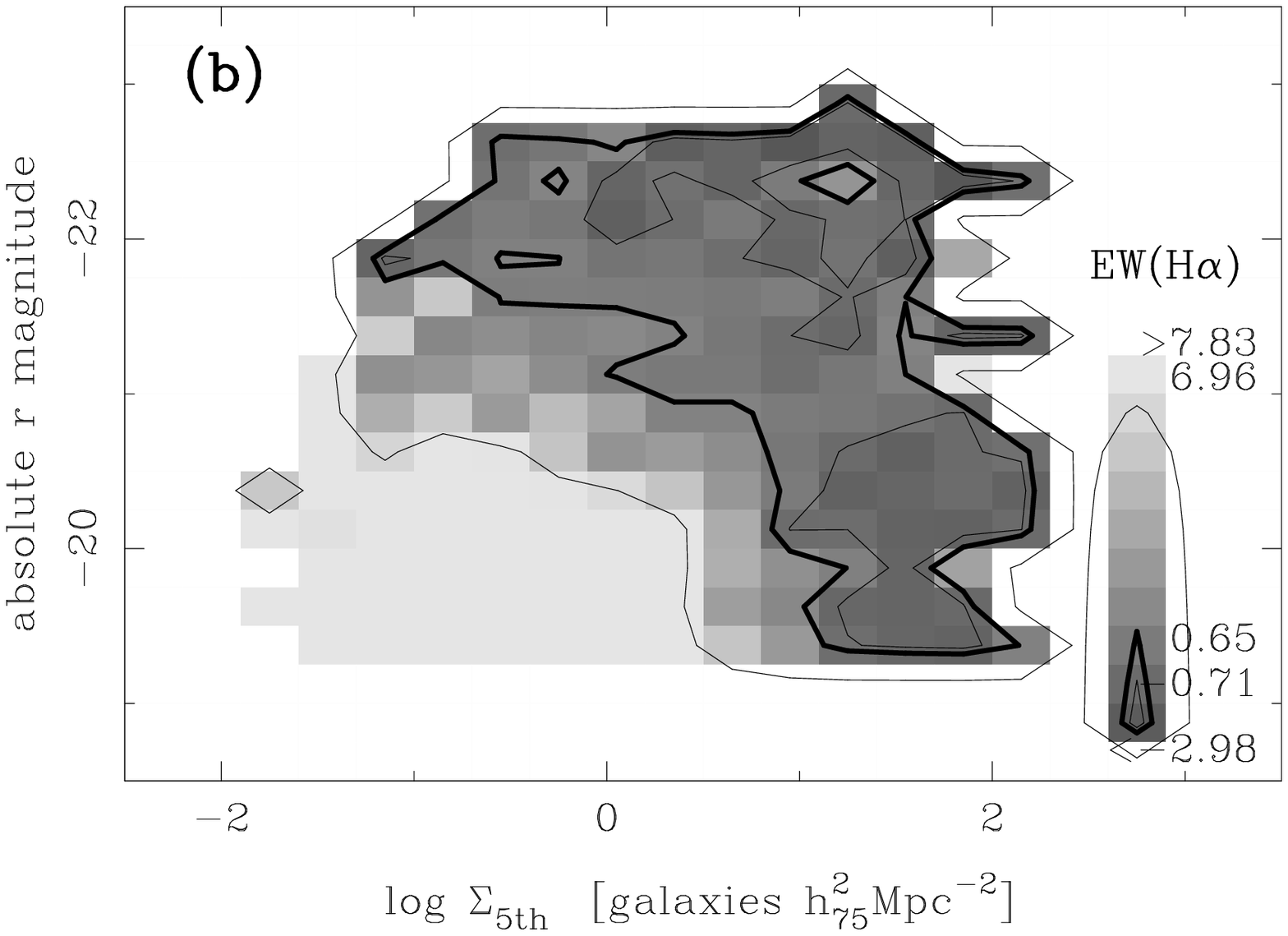}
\plottwo{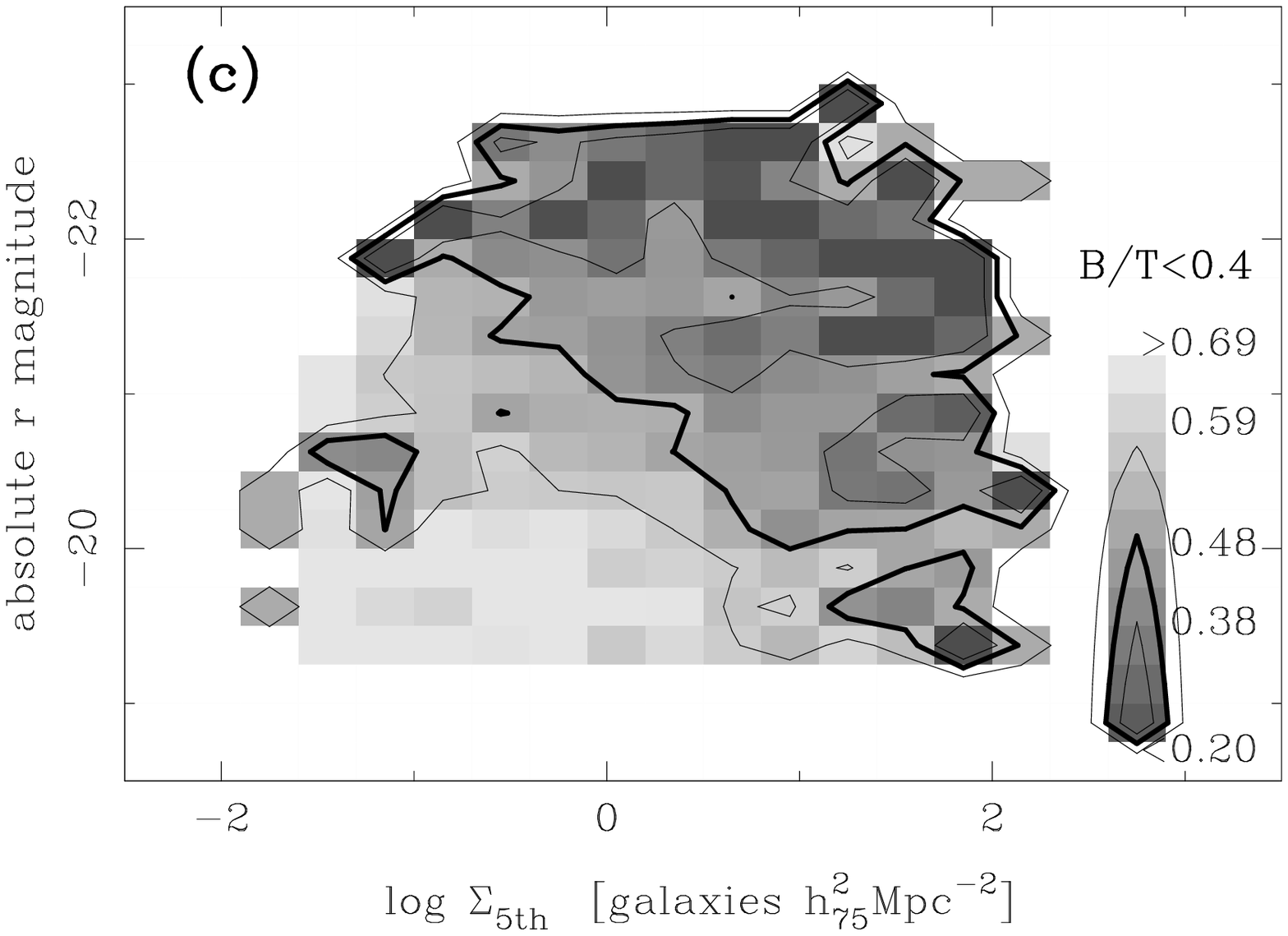}{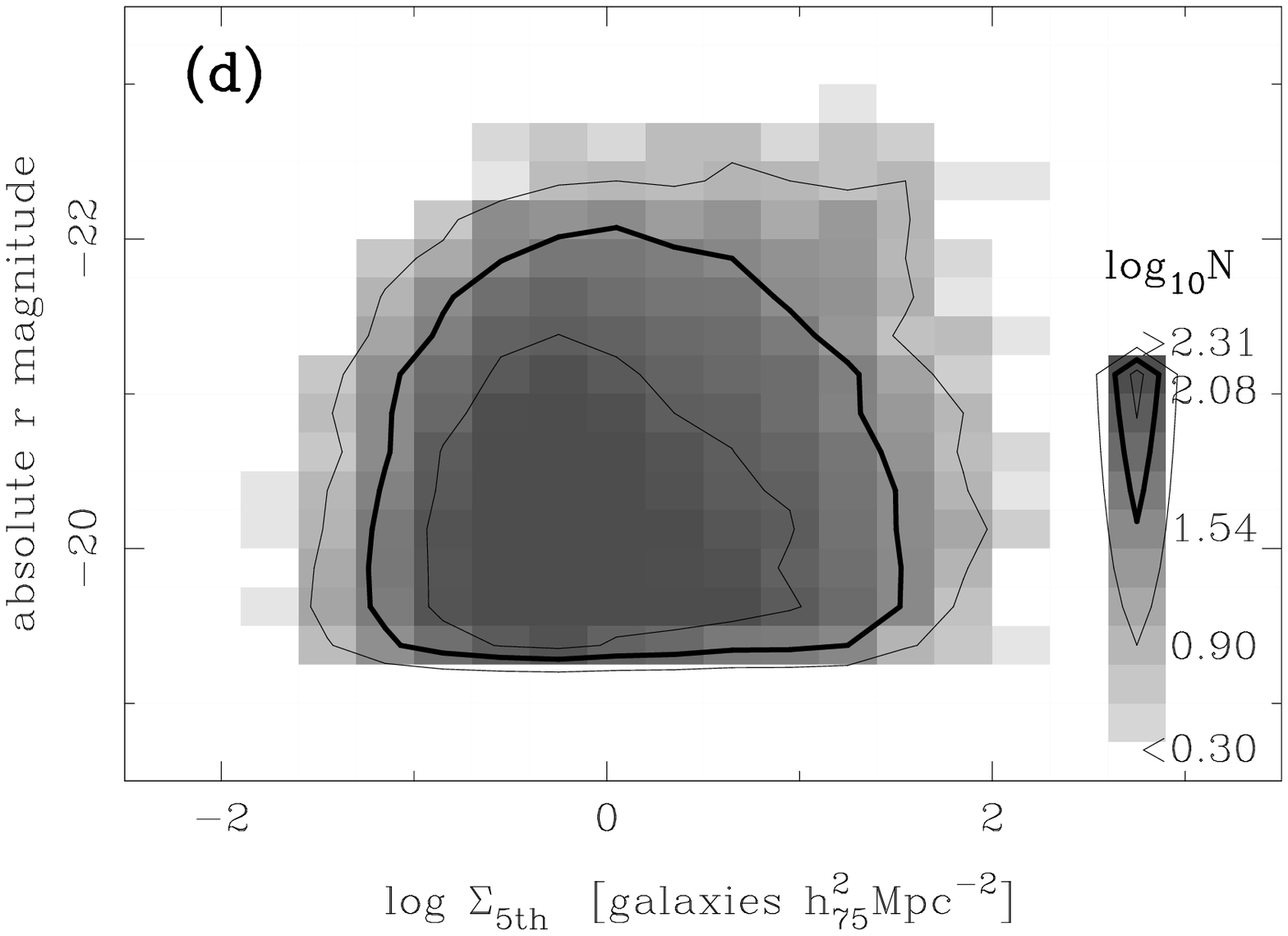}
\caption{
Distribution of {\bf (a)} $g-i$, {\bf (b)} EW(H$\alpha$), {\bf (c)} \btn,
and {\bf (d)} number of galaxies on the absolute magnitude and local density plane.
{\bf (a)}
The brightness of each grid represents the median $g-i$ of galaxies in the grid.
Darker grids represent redder $g-i$ colors.
It should be noted that the number of galaxies varies from grid to grid,
and thus the statistical significance differs from grid to grid.
Panel (d) shows galaxy counts in each grid.
The contours are the median and quartiles of the colors of the grids.
{\bf (b)}
Same as (a), but for EW(H$\alpha$).
Brighter grids mean more active star formation.
{\bf (c)}
Same as (a), but for the late-type fraction ($B/T<0.4$).
Brighter grids mean a larger fraction of late-type galaxies.
{\bf (d)}
Galaxy counts in the logarithmic scale. Darker grids mean higher counts.
\label{fig:local_vs_mag_sf}
}
\end{figure}

Further illustrations of the relationship between star formation, morphology,
luminosity, and local density are presented in this appendix.
Figure \ref{fig:local_vs_mag_sf} shows the dependence of star formation and morphology
on the local density and absolute magnitude plane.
As found in Figures \ref{fig:local_vs_color} and \ref{fig:local_vs_ha},
star formation is a strong function of both local density and luminosity.
In very dense regions ($\log \Sigma_{\rm 5th} \gtrsim 1$), there is no strong correlation with either
local density or luminosity, and most of the galaxies are red.
The strong luminosity dependence can be seen in the field region ($\log \Sigma_{\rm crit} \lesssim$ 0.4),
in the sense that fainter galaxies are more actively forming stars.
It should be noted that the strong aperture bias should contribute to the luminosity dependence (see Appendix B).
However, the fact that $g-i$ and \ewha show the identical luminosity dependence
suggests that the contribution is small.
The trends seen for morphology are similar to those found for star formation,
although the luminosity dependence seems to be stronger.

\section{Correlations between Environmental Parameters}

\begin{figure}
\epsscale{1}
\plottwo{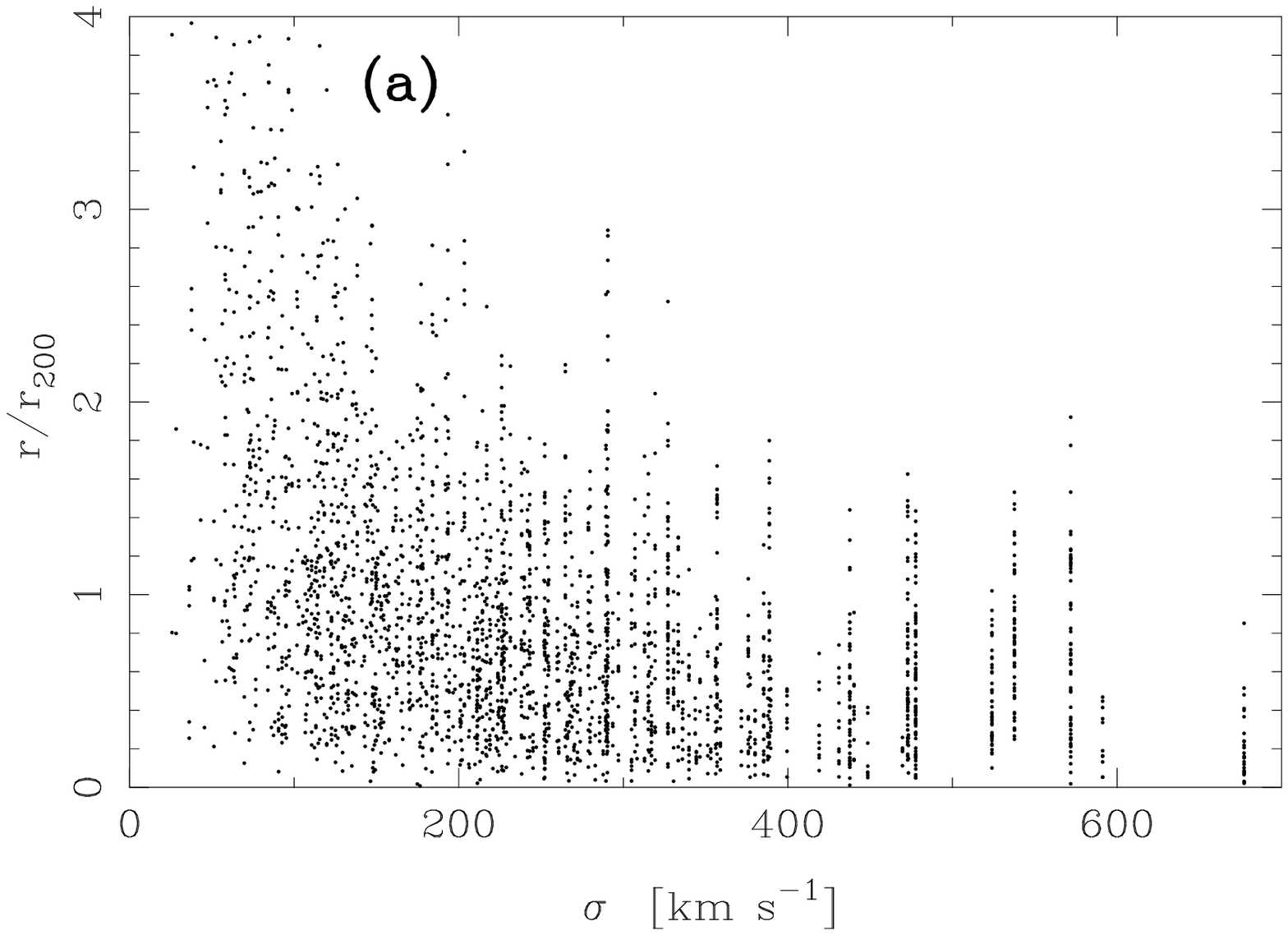}{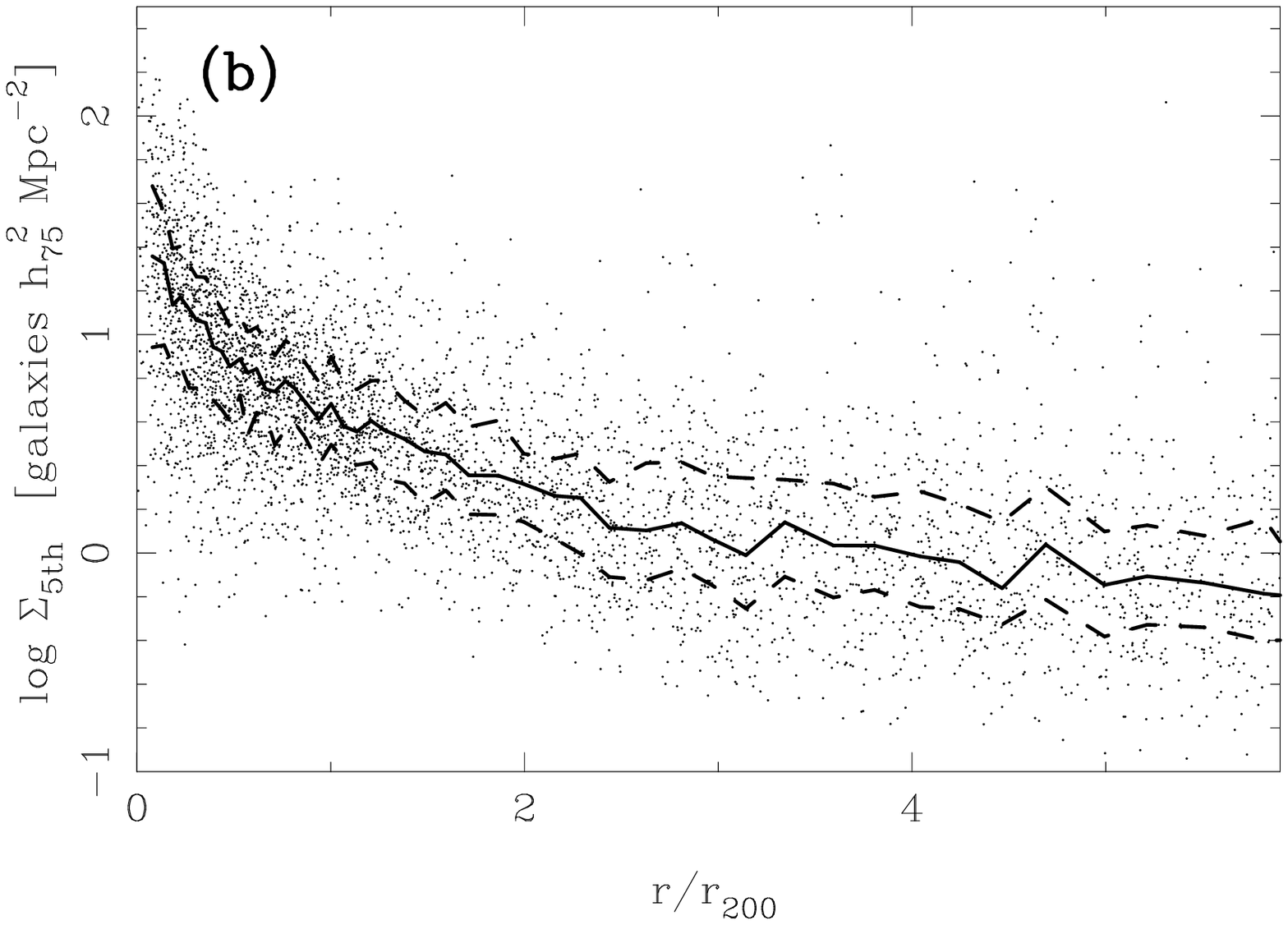}
\caption{
Some representative correlations between various environmental parameters.
{\bf (a)} Projected system-centric distance as a function of velocity dispersion.
{\bf (b)} Local density plotted against the projected distance from the nearby system center.
Galaxies whose redshifts are within $3\sigma$ of the nearby system redshift are plotted.
The solid and dashed lines show the median and the quartiles of the distribution.
\label{fig:env_corr}
}
\end{figure}

We show some representative examples of the correlation
between environmental parameters in Figure \ref{fig:env_corr}.
Figure \ref{fig:env_corr}(a) plots the radial distribution of galaxies in the ``linked list'' of the FOFA
as a function of velocity dispersion.
The projected center of a system is estimated by averaging the projected galaxy positions weighted by the
velocity difference with respect to the system velocity center.
We use $r_{200}$ as an estimate of the virial radius of a system.
The $r_{200}$ is defined as the radius in which the mean interior density is 200 times the critical density
of the universe.
In our cosmology, $r_{200}$ can be obtained from
\begin{equation}
r_{200}=2.3\sigma (1+z)^{-1.5}\ \ h^{-1}_{75}\rm\ kpc,
\end{equation}
where $\sigma$ is velocity dispersion of a system in unit of kilometers per second \citep{carlberg97}.
It can be seen that galaxies in the FOFA systems extend beyond the virial radius ($\sim1.5r_{200}$).
That is, our sample covers the outskirts of galaxy systems.
Galaxies in poor systems extend far beyond the virial radius.
This might be because galaxies in such poor systems are not yet virialized, and the assumption of
the virial theorem in the calculation of $r_{200}$ is not valid.

Figure \ref{fig:env_corr}(b) plots local density against clustercentric radius.
In this panel, only galaxies within $3\sigma$ from the system redshift are plotted.
There is a clear correlation between the two quantities.
We have reported that there is the critical density at $\log\Sigma_{\rm crit}\sim 0.4$,
at which point galaxy properties abruptly change.
It can be seen that the critical density corresponds to $1-2\ r_{200}$.

\clearpage

\clearpage

\clearpage

\clearpage

\clearpage

\clearpage

\clearpage

\clearpage

\clearpage

\clearpage

\clearpage

\clearpage

\clearpage

\clearpage

\clearpage

\clearpage
\begin{deluxetable}{crccc}
\tabletypesize{\scriptsize}
\tablecaption{Definition of the two subsamples\label{tbl-1}}
\tablewidth{0pt}
\tablehead{
\colhead{Subsample} & \colhead{Definition} & \colhead{Total number} &
\colhead{Local density estimated} & \colhead{non-AGNs\tablenotemark{a}}
}
\startdata
Bright & $M_r<M_r^*+1$ & $8,794$ & $5,599$ & $4,894$\\
Faint & $M_r^*+1<M_r<M_r^*+2$ & $10,920$ & $6,777$ & $6,108$
\enddata
\tablenotetext{a}{Number of galaxies that have local density estimates and are classified as non-AGNs.}
\end{deluxetable}

\clearpage
\begin{deluxetable}{crcc}
\tabletypesize{\scriptsize}
\tablecaption{Dependence of galaxy properties on surface galaxy density \label{tbl-3}}
\tablewidth{0pt}
\tablehead{
\colhead{Subsample} & \colhead{Definition} &
\colhead{Trends of star formation ($g-i$ and \ewha)} &
\colhead{Trends of morphology (\btn)}
}
\startdata
Brightest & $M_r<M_r^*-1$ & no correlation with density & no correlation with density\\
Bright & $M_r<M_r^*+1$ & no break, monotonic change with density & no break, monotonic change with density\\
Faint & $M_r^*+1<M_r<M_r^*+2$ & break at $\log\Sigma_{\rm crit}\sim 0.4$ &
break at $\log\Sigma_{\rm crit}\sim 0.4$\\
\enddata
\end{deluxetable}

\clearpage
\begin{deluxetable}{crcc}
\tabletypesize{\scriptsize}
\tablecaption{Dependence of galaxy properties on system richness \label{tbl-4}}
\tablewidth{0pt}
\tablehead{
\colhead{Subsample} & \colhead{Definition} &
\colhead{Trends of star formation and morphology (\ewha and \btn)}\\
}
\startdata
Bright & $M_r<M_r^*+1$ & no clear correlation with $\sigma$\\
Faint & $M_r^*+1<M_r<M_r^*+2$ & dominated by non-star-forming galaxies in $\sigma>200\kms$ systems\\
\enddata
\end{deluxetable}


\end{document}